\documentclass[preprint]{aastex}

\newcommand{\subsun}{\mbox{$_{\odot}$}}
\newcommand{\etal}{{\it et al.\/}}
\newcommand{\teff}{$T_{eff}$}
\newcommand{\grav}{log($g$)}

\newcommand{\eqw}{$W_{\lambda}$}

\begin{document}

\title{The Chemical Evolution of the Ursa Minor Dwarf Spheroidal Galaxy\altaffilmark{1} }

\author{Judith G. Cohen\altaffilmark{2} and Wenjin Huang\altaffilmark{3} }

\altaffiltext{1}{Based on observations obtained at the
W.M. Keck Observatory, which is operated jointly by the California
Institute of Technology, the University of California and the National
Aeronautics and Space Administration}

\altaffiltext{2}{Palomar Observatory, Mail Stop 105-24,
California Institute of Technology, Pasadena, California 91125
(jlc@astro.caltech.edu)}

\altaffiltext{2}{Palomar Observatory, current
address: University of Washington, Astronomy, Box 351580, Seattle,
Washington, 98195 (hwenjin@astro.washington.edu)}

\begin{abstract}

We present an abundance analysis based on high resolution
spectra of 10 stars selected to span the full range in metallicity
 in the Ursa Minor dwarf spheroidal galaxy.  We find
[Fe/H] for the sample stars ranges from $-1.35$ to $-3.10$~dex.
Combining our sample with previously published work for a total
of 16 luminous UMi giants, we establish the  trends of
abundance ratios [X/Fe] as  functions of [Fe/H]
for 15 elements. In key cases, particularly for the
$\alpha$-elements, these trends resemble
those for stars in the outer part of the Galactic halo, especially
at the lowest metallicities probed.
The neutron capture elements show a $r$-process distribution
over the full range of Fe-metallicity reached in this dSph galaxy.
This suggests that the duration 
of star formation in the UMi dSph was shorter than in 
other dSph galaxies.
The derived ages for a larger sample of UMi stars with
more uncertain metallicities also suggest a  population dominated by
uniformly old ($\sim$13~Gyr) stars, with a hint
of an age-metallicity relationship.

In comparing our results for UMi, our earlier work in Draco, and
published studies of more metal-rich dSph Galactic satellites,
there appears to be a pattern of moving from
a chemical inventory for dSph giants with [Fe/H] $\lesssim -2$~dex
which is very similar to that of stars in the
outer part of the Galactic halo (enhanced 
$\alpha$/Fe relative to  the Sun,
coupled with subsolar [X/Fe] for
the heavy neutron capture elements and $r$-process domination),
switching to subsolar $\alpha$-elements
and super-solar  $s$-process dominated neutron capture elements
for the highest [Fe/H] dSph stars.  The combination of
low star formation rates over a varying and sometimes
extended duration that produced the stellar
populations in the local dSph galaxies with [Fe/H] $> -1.5$~dex
leads to a chemical inventory wildly discrepant from that of  any
component of the Milky Way.

We note the presence of two UMi giants with [Fe/H] $< -3.0$~dex in our sample,
and reaffirm that the inner Galactic halo could have been formed
by early accretion of Galactic satellite galaxies and dissolution
of young globular clusters, while the outer halo could have formed
from those satellite galaxies accreted somewhat later.

\end{abstract}

\keywords{galaxies: individual (UMi), galaxies: abundances, galaxies: dwarf}

\section{Introduction\label{intro}}

The Ursa Minor (UMi) dwarf spheroidal (dSph) galaxy is a 
satellite of the Milky Way at a distance of about
70~kpc \citep{mighell99}.  It is the least luminous of the
8 classical dSph satellites, with $L_V = 3 \times 10^5~L_{\odot}$
\citep{grebel03}.
\cite{cos_umi} provide a proper motion survey
and photometry for this galaxy down to the level of the 
horizontal branch.  With $\sim$450 members, they 
found that the stellar population of the UMi dSph
resembles that of an old metal-poor Galactic globular cluster
with a steep red giant branch (RGB) and a blue horizontal branch.  The HST
imaging study of \cite{mighell99} confirms this simple star formation history
for UMi, suggesting a single major burst of star formation 
about 14~Gyr ago that lasted less than 2~Gyr.  
The photometric survey by \cite{bellazzini02}
established a mean abundance [Fe/H] $-1.8$~dex
with a spread
of $\sim$0.5~dex within UMi, assuming all stars are of a similar age.
There is no evidence for multiple main sequences nor any sign
of ongoing star formation.  Very low
upper limits were established in UMi for neutral hydrogen gas
by \cite{young00} and for ionized H by  \cite{gallagher03}.

Many radial velocity surveys \citep[see, e.g.][]{cos_umi,palma03}
of large samples of stars have been carried out to
determine membership and 
measure the stellar velocity dispersion as a function
of radius in UMi,
while \cite{walker07} highlights the very large
$v_r$ datasets that
can be assembled today for such galaxies.
The observed $\sigma_v$ in UMi, as is also true for Draco,  is 
unexpectedly high, given the
low luminosity of the system, and remains flat to large radii,
which according to \cite{penarrubia08} (see also the references therein)
requires the presence of large
amounts of dark matter.
These studies generally ignore the issue of potential ongoing 
tidal disruption
affecting the internal kinematics of the Milky Way satellite.
\cite{piatek01} \citep[see also][]{bellazzini02,palma03} 
review the evidence for substructure in UMi, which
might be an argument for tidal
effects..

There is great current interest in the detailed
properties of satellites of the Galaxy in light of our
greatly improved hierarchical cold dark matter cosmological models, which
gave rise to the the missing
satellite  problem \citep{klypin99}.  This is only enhanced
by the  discovery that the dSph galaxies appear to be dark matter dominated systems,
unlike globular clusters of similar total stellar mass.
With the advent of 
efficient high dispersion
spectrographs, large area CCD detectors, and 10-m class telescopes,
studying stars in at least the nearer Galactic satellites
at high spectral dispersion has become feasible. 

High resolution spectroscopy was obtained for 6 UMi giants
by \cite{shetrone2}, and for three (including two duplicates from
the earlier work) by \cite{subaru}, one of which was reexamined
in an attempt to detect the radioactive actinide thorium by
\cite{aoki08}.
In this paper we present detailed
abundance analyses of a sample of 10 luminous UMi stars near
the RGB tip, which more than doubles the sample of UMi stars from
the earlier works. 
Our goal is understanding the chemical evolution
of UMi, and how this and other dSph galaxies may
be related to the population of
Galactic halo field stars and to Galactic globular
clusters.   The sample is presented in \S\ref{section_param}
where the procedure for determining their stellar parameters is described.
The next section describes the observations, while \S\ref{section_analysis}
gives some details of the abundance analysis.
We compare
our results to those for Galactic
halo field stars in \S\ref{section_field}, apply our
toy model for abundance ratios in \S\ref{section_toy_model},
and discuss the age-metallicity relation in UMi in \S\ref{section_age}.
A discussion of the predictions of nucleosynthesis  
and  chemical evolution models
as applied to UMi and to Draco is given
in \S\ref{section_chemev}. We speculate on the 
role the dSph satellite galaxies might have played the
formation of the Galactic halo in \S\ref{section_haloform}. 
A brief summary concludes the paper.

This paper is a sequel to our earlier study of the chemical evolution of the
Draco dSph \citep{cohen_draco} (henceforth C09).
The techniques used are similar and the reader is urged to consult
our earlier work for additional details as necessary.

\section{Stellar Sample and Stellar Parameters \label{section_param}}

Our sample in the UMi dSph galaxy contains 10 stars; details
are given in Table~\ref{table_sample}.
It was selected from Table~3.6 of \cite{winnick03} to include
stars which are known radial velocity members of this satellite
to the Galaxy
at or near the RGB tip spanning the full range in color 
and in metallicity, and not previously observed at high spectral
resolution.
Winnick measured the infrared Ca triplet in moderate
resolution spectra of a sample of UMi stars 
chosen from earlier radial velocity surveys,
excluding
known carbon stars \citep[see, e.g.][]{shetrone1}.
Details of her calibration with metallicity and related
issues were discussed in the appendix to C09.

We adopt the procedures described in \cite{cohen02} and used in all
subsequent work by the first author 
published to date to determine the stellar parameters
for our sample of luminous UMi giants.
Our \teff\ determinations are based on the broad band colors
$V-I$, $V-J$ and $V-K$ and the predicted colors from the model
grid of \citet{houdashelt00}.
The optical photometry is 
from the SDSS \citep{york00} using the transformation equations of
\cite{smith02}.
The IR photometry is taken from 2MASS \citep{2mass1,2mass2}, and is
transformed from the 2MASS system to the Johnson-Bessell system
using the results of \cite{carpenter}.
The galactic extinction is from the map of 
\cite{extinct98}; E($B-V$) does not exceed $0.03$~mag for any star in the 
UMi sample.

We derive surface gravities by combining these \teff\ 
with bolometric corrections from the model grid,
the observed $V$ corrected for reddening, an assumed
stellar mass of 0.8~$M_{\odot}$, and the distance to the
UMi dSph galaxy.   We use the [Fe/H]
values of \cite{winnick03} from the infrared Ca triplet as an 
initial guess.
We iterate as necessary given the metallicity we derive here through
analysis of our high resolution spectra\footnote{This was also the procedure
used for the Draco HIRES sample; the description of the procedure
for determinations
of \grav\ given in C09 is not correct.}.

The resulting stellar parameters, which have been
derived with no reference to the spectra themselves,
are given in the second and third columns of
Table~\ref{table_param}, as are the
heliocentric radial velocities.  The random uncertainties
in the adopted \teff\ from photometric errors
are 100~K.  This ignores systematic errors
which may be present.  The adopted uncertainties in \grav\ 
based on the uncertainties in \teff, the stellar mass,
and the distance to UMi are 0.2~dex.

\begin{figure}
\epsscale{0.8}
\plotone{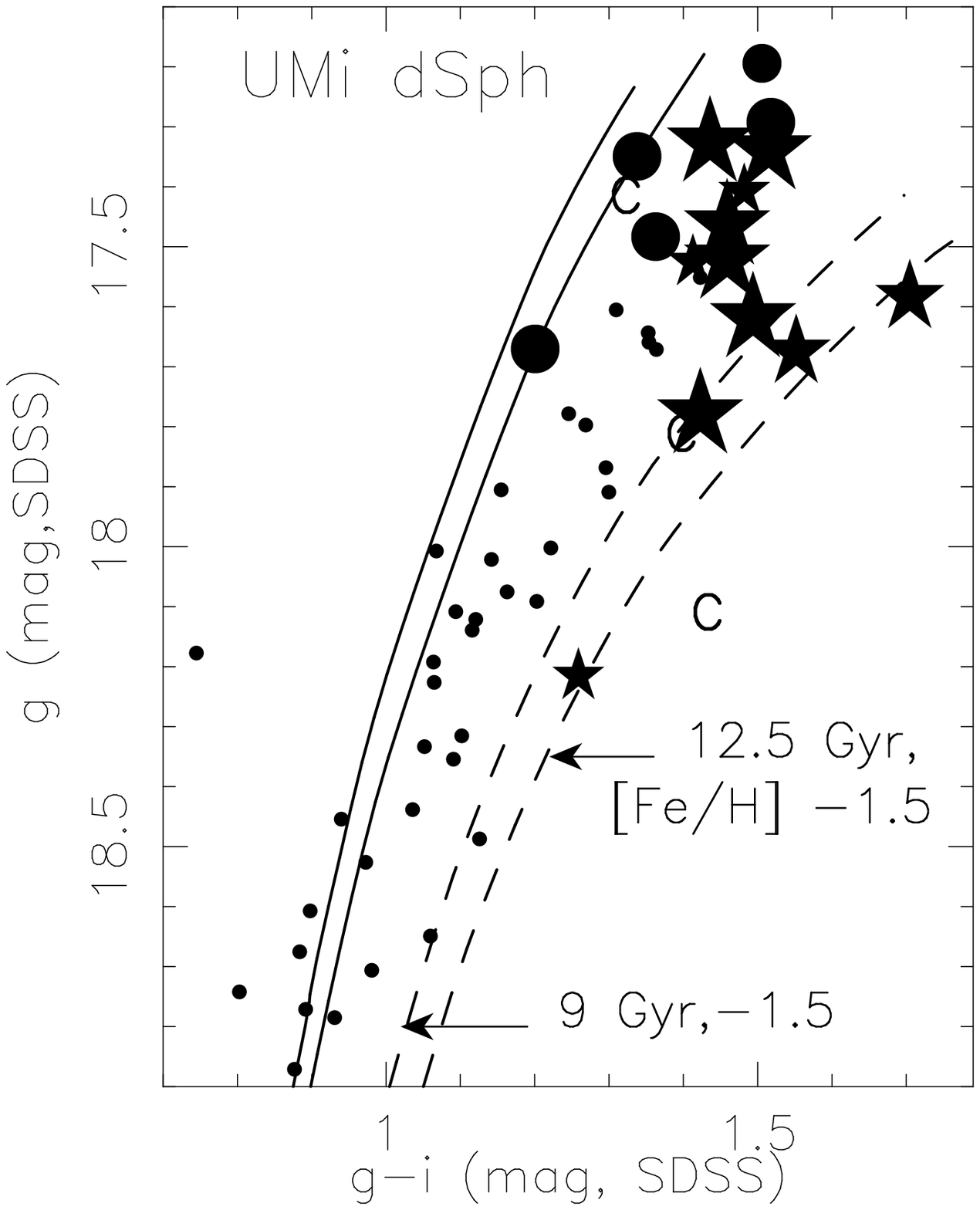}
\caption[]{Our UMi HIRES sample is shown in a plot of
$g'-i$ versus $g'$  (large symbols),
where filled circles indicate giants with [Fe/H] $< -2.5$~dex, and
star symbols have higher Fe-metallicity.
The sample of UMi stars studied by \cite{shetrone2} and by \cite{subaru}
are indicated by the small and intermediate sized symbols.
The dots indicate UMi members from \cite{winnick03} with  photometry
from the SDSS.  
Carbon stars that are confirmed members of UMi from \cite{shetrone1}
are indicated by the letter C.  All observerational data are  corrected for 
interstellar reddening.
Isochrones from the Dartmouth Stellar Evolution Database
\citep{dotter08} for [Fe/H] $-2.5$~dex with [$\alpha$/Fe] = +0.2~dex
(solid lines)
and for [Fe/H] $-1.5$~dex with [$\alpha$/Fe] Solar (dashed lines)
for ages 9 and 12.5 Gyr are shown.
\label{figure_isochrone}}
\end{figure}

Fig.~\ref{figure_isochrone} shows 
our sample of 10 luminous giants in UMi in a plot of
$g'-i'$
versus $g'$  corrected for interstellar reddening;
the previously studied sample of 6 giants
from \cite{shetrone2} and by \cite{subaru} is shown as well.
Members of UMi from the list of \cite{winnick03} were cross indexed
with the SDSS photometry \citep{york00} from DR7 \citep{sdss_dr7}
and are also displayed.
Superposed in this figure are isochrones from the Dartmouth
Stellar Evolution Database \citep{dotter08} for [Fe/H] 
 $-2.5$~dex with [$\alpha$/Fe] = +0.2~dex
(solid lines) and for [Fe/H] $-1.5$~dex with [$\alpha$/Fe] Solar (dashed lines)
for ages 9 and 12.5 Gyr.

Our HIRES sample of luminous UMi giants was selected to span
the full range in metallicity as inferred from 
the Ca triplet indices by \cite{winnick03}.  Fig.~\ref{figure_isochrone} shows
it does cover
the full range in $g'-i'$ color of the upper RGB of UMi members.
The luminosity of the brightest UMi giants is in good agreement
with that predicted from the isochrones for the RGB tip as a
function of metallicity in the $g',i'$ colors.  The
deduced ages will be described later in \S\ref{section_age}.

\section{Observations \label{section_obs} }

The UMi stars in our sample were observed with HIRES-R \citep{vogt94} at 
the Keck~I telescope
during three runs, in June 2008, Aug. 2009, and Feb. 2010. 
Sky conditions were good during all of these runs. 
An earlier run
in 2005 was assigned for this purpose, but no usable spectra could
be obtained at that time.
The instrument configuration yielded complete spectral coverage
in a single exposure from 3810 to 6700~\AA, and extends
to 8350~\AA\ with small gaps between orders.
The slit width was  1.1 arcsec
($\lambda/\Delta\lambda = 35,000$) for all exposures.
The total
exposure times for each star are given in Table~\ref{table_sample};
the exposures were broken 
up into 1800 or 2400 sec segments to expedite
removal of cosmic rays.
The signal-to-noise ratios (S/N)
per spectral resolution element at 5800~\AA\ in the continuum
near the center of the echelle order are given in the
last column of this table; they range from 80 to 100,
but the S/N drops towards the bluer part of the spectra, becoming poor 
(less than 40) at the
bluest end of these spectra.
This S/N calculation utilizes only
Poisson statistics, ignoring issues of cosmic ray removal,
night sky subtraction, flattening, etc.   

The processing of the spectra was done with 
MAKEE\footnote{MAKEE was developed
by T.A. Barlow specifically for reduction of Keck HIRES data.  It is
freely available on the world wide web at the
Keck Observatory home page, 
http:\\www2.keck.hawaii.edu/inst/hires/data\_reduction.html.} and Figaro 
\citep{shortridge93} 
scripts, and follows closely that described by \cite{cohen06}.
The equivalent widths were measured as described in \cite{cohen04}.
Due to the lower S/N in the blue, lines bluer than 4400~\AA\
were ignored if the species had sufficient other 
detected lines.
Lines with \eqw $> 175$~m\AA\ were
discarded except for two lines from the Mg triplet,
the Na~D lines and Ba~II lines in some of the stars; for these
key elements no or only a few weaker features could be detected
in most of the stars.  
Table~\ref{table_eqw} lists the atomic parameters adopted
for each line and their equivalent widths measured in the spectra of
each of the UMi dSph stars. 

UMi~N37, 33533, and 41065 show
strong emission in the blue wing
of H$\alpha$ in their spectrum; the first two also show
weaker emission in the red wing of this line.
UMi~COS233 \citep[COS ID numbers are from][]{cos_umi}
and JI2 show weak emission in the blue
wing of H$\alpha$, with no emission detected in the red wing.
No other anomalies were noted from visual inspection of the spectra.

\section{Analysis \label{section_analysis} }

The analysis is identical to that of \cite{cohen08}
and earlier references therein.
In particular we use the model
stellar atmosphere grid of \cite{kurucz93}
and a current
version of the LTE spectral synthesis program MOOG \citep{moog},
which treats scattering as LTE absorption.

Our analysis assumes classical plane
parallel stellar atmospheres and LTE, both for atomic
and for molecular features.
We adopt a Solar Fe abundance of log[$\epsilon$(Fe)] = 7.45~dex
based on our solar spectrum analysis, see also
\cite{asplund05}.
This value is somewhat lower (by up to 0.10~dex) than that used by 
many groups, which  
leads directly to our [Fe/H] values for a given star being 
somewhat higher  and to our abundance 
ratios [X/Fe] being somewhat lower than those which would be inferred by 
most other teams.  
Our $gf$ values are generally taken from
Version 3.1.0 of the NIST Atomic Spectra Database 
(phsics.nist.gov/PhysRefData/ASD/index.html, NIST Standard Reference
Database 78).
A comparison of $\log gf$ values for Fe we adopt 
with those of the First Stars Project at the VLT \citep{cayrel04}
was given in \cite{cohen08}, and shows excellent agreement
for both Fe~I and Fe~II.
Corrections for hyperfine structure for
Sc~II, V~I, Mn~I, Co~I, Cu~I, Ba~II, and Eu~II
were used when necessary: the majority of the HFS patterns were adopted from
\cite{prochaska}.

Our abundances for C are from the 4320~\AA\ region of the
G band of CH, where the absorption is less than in
the main part of the G band at 4300~\AA.  O abundances
are primarily from the forbidden line at 6363~\AA; 
the radial velocity of UMi often shifts the 6300~\AA\ [O~I] line
to overlap the strong terrestrial atmospheric line at 6295.2~\AA, making
it not usable.
Our nominal Solar C and O abundances are 8.59 and 8.83~dex respectively.
See C09 for further comments on the molecular abundances.

%

Since the UMi stars are rather faint for 2MASS, the uncertainties
in the $K_s$ magnitudes are fairly large, ranging up to 0.09~mag.  
We therefore
feel free to
slightly adjust \teff\ and \grav\  after the first pass
through the analysis to improve the ionization equilibrium and
slope of the abundances determined from the set of Fe~I lines
as a function of $\chi$ (the excitation potential of the lower level).
These spectroscopic stellar parameters are given in the
fourth and fifth column of Table~\ref{table_param} and
are the ones used subsequently.  With these values we were
able to achieve good ionization equilibrium for Ti and Fe as
well reasonable  excitation equilibrium of Fe~I.
Table~\ref{table_slopes} gives the slope of a linear fit
to the abundances determined from the set of Fe~I lines
as a function of $\chi$,
\eqw, and $\lambda$, which are most sensitive to \teff, $v_t$,
and the wavelength dependence of any problems in establishing the correct
location of the continuum (perhaps arising from the more severe
crowding towards bluer wavelengths) or of a
missing major source
of continuous opacity, respectively.  There are $\sim$ 75 to 190 Fe~I lines
detected in
each star, with $\chi$ ranging from 0 to $\sim$4.5~eV.  The only
slope which is large enough to be of concern
and which tends to have the most significant correlation coefficient 
is that with $\chi$ ($\mid cc(\chi) \mid ~ > ~ 0.4$ for some of the sample giants),
which depends largely on \teff.   In our final adopted
solutions, the Fe~I slope as a function of $\chi$
tends to be slightly negative, with values ranging from
$-0.02$ to $-0.11$~dex/eV, with small $\mid cc(\chi) \mid$ for
the most negative values.  This slope
decreases by $\sim$0.1~dex/eV/($\Delta$\teff = +250~K). 
A decrease in \teff\
of a maximum of 125~K, consistent with our adopted \teff\ uncertainty,
would make all these slopes consistent with zero, and would decrease
the [Fe/H] derived from Fe~I lines by $\sim$0.2~dex, but would ruin the
ionization equilibrium of Ti.

One potential concern is the possibility of non-LTE in Fe affecting
the ionization equilibrium.
This is discussed in detail in C09.
The slightly negative Fe~I slope with excitation
potential mentioned above
may be a sign that overionization of Fe is occurring.  If this were the
case,  we would  have been driven to adopt a higher \teff\
than the actual value;
our derived [Fe/H] values would be too high as indicated above,
but the deduced abundance ratios would not be significantly
affected by such a decrease in \teff. With this in mind, we adopt 
asymmetrical uncertainties for \teff\ of +100~K, $-150$~K.
Since we have been
able to achieve satisfactory ionization equilibrium for Fe and for Ti
and at the same time
reasonably good excitation equilibrium for Fe 
with a single value of \teff\ which 
differs from that set solely from broad band photometry by 50~K or
less for more than half of the UMi giants, we regard our choices for stellar
parameters  as satisfactory.  
Ideally, of course, one would like to have a full non-LTE 3D
analysis including both convection and spherical (as distinct
from plane parallel) layers for all species,  but at the present time this is 
not practical.

Our derived abundances for the 10 UMi luminous giants are given in 
Tables~\ref{table_abunda} and \ref{table_abundb}.  The sensitivity
of the absolute and relative abundances for each species detected
to small changes in 
\teff, \grav, 
microturbulent velocity, and assumed [Fe/H] for the stellar atmosphere
model are similar to those we calculated for  Draco; see Tables~5 and 6
of C09.  The only non-LTE correction we have made is to the
Al abundance when the 3961~\AA\ resonance line of Al~I was used;
in many cases this was the
only feature of Al that could be detected.
We adopt a correction of +0.60~dex  based on the
calculations of \cite{bau96,bau97}. 

\begin{figure}
\epsscale{0.9}
\plotone{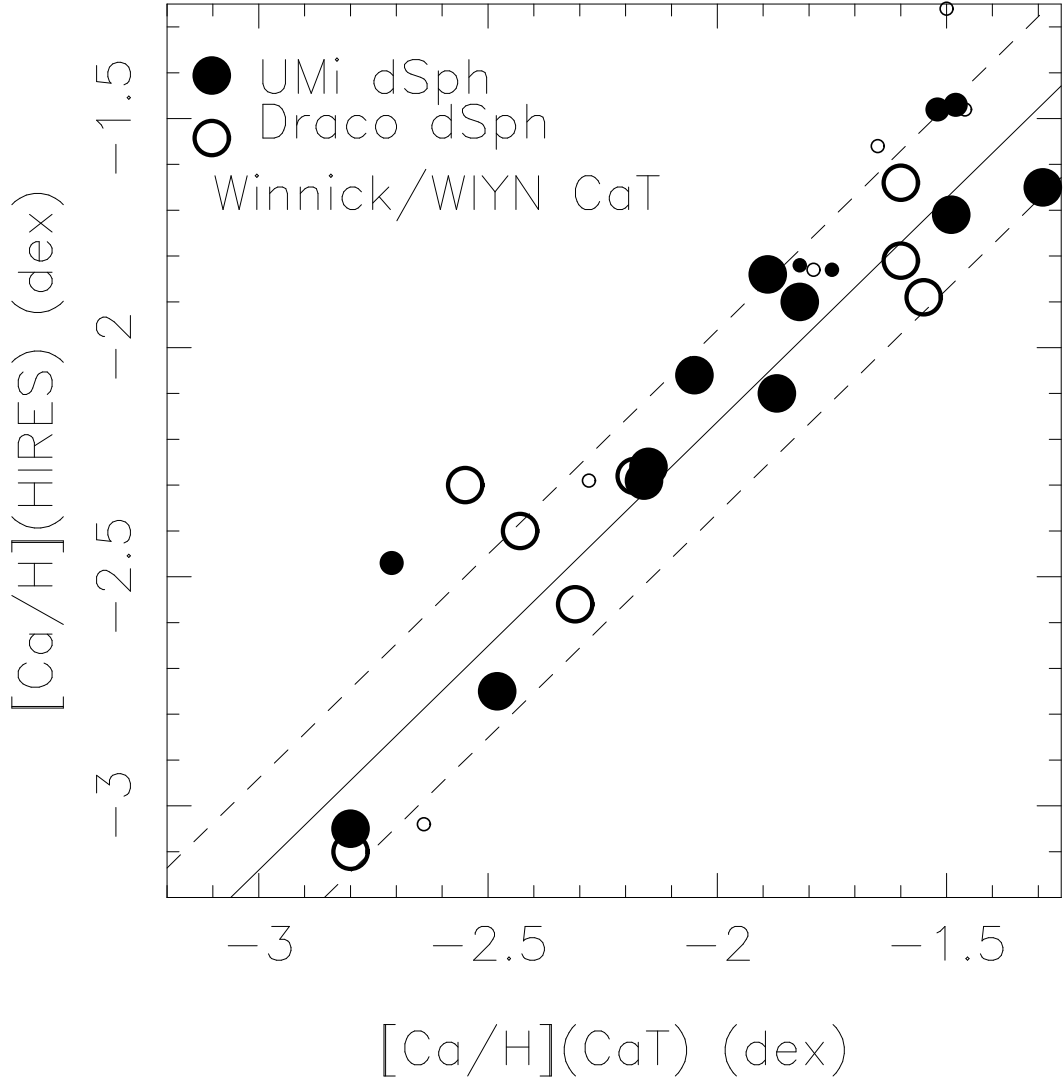}
\caption[]{The [Ca/H] values derived
from indices of the strength of
the infrared Ca triplet by \cite{winnick03} are compared to
the results of our high-resolution detailed abundance analyses
for our sample of 10 RGB stars in each of the UMi and the Draco dSph galaxy.
The solid line
represents the best fit to the UMi data, while the dashed lines show offsets between
the two determinations of $\pm 0.2$~dex.  Smaller symbols denote 3 UMi
stars from \cite{subaru}, while the smallest symbols are from \cite{shetrone2}.
\label{figure_winnick}}
\end{figure}

We compare the [Ca/H] derived by \cite{winnick03} based on
her infrared Ca triplet indices with our values from HIRES spectra.
The result for UMi and for Draco from C09 is shown in Fig.~\ref{figure_winnick}.  
We find (for UMi only) that 
[Ca/H](HIRES) $~= ~-0.21~+ ~0.98~ \times ~$ [Ca/H](Winnick/CaT] 
with $\sigma$ around the linear fit of only 0.13~dex.

\section{Comparison with Galactic Halo Field Stars \label{section_field}}

We compare the behavior of abundance ratios
within UMi  to those of Galactic halo field stars
in detail. The sample of UMi stars with detailed abundance analyses
based on high dispersion spectra is now 16, including the 10
we present here.   \cite{shetrone2}
presented an analysis for 6 UMi members; there is no overlap
with our sample.
We ignore their star K, which they state is a carbon star.  
Better spectra taken
with HDS at the Subaru telescope were analyzed by
\cite{subaru} for three stars, two of which were included in
\cite{shetrone2}. 
In view of
the much higher S/N of the \cite{subaru} spectra, we adopt
their abundances for these two 
stars\footnote{Our spectra have S/N considerably
higher than those of \cite{shetrone2}, and perhaps
slightly lower than those of  \cite{subaru}.}.

\begin{figure}
\epsscale{0.9}
\plotone{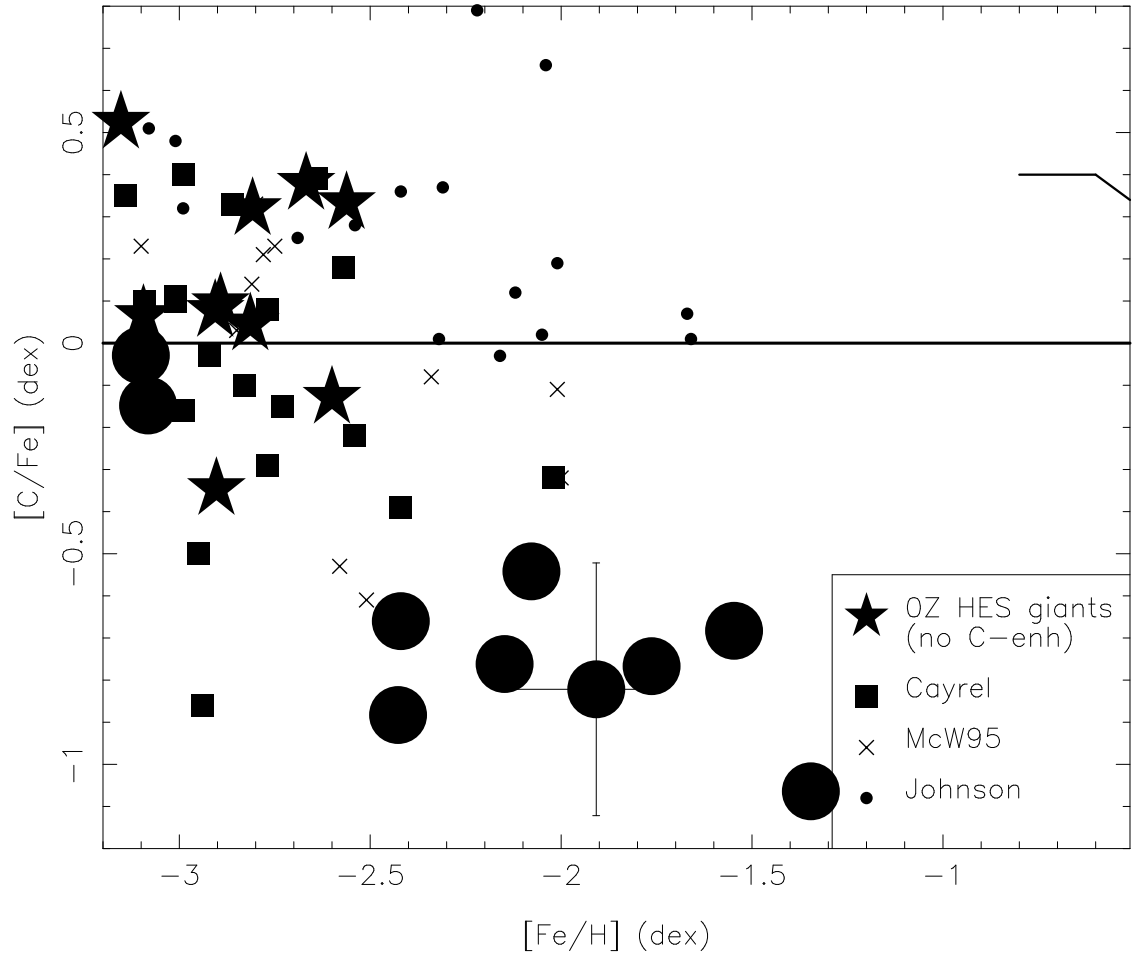}
\caption[]{[C/Fe] from the G band of CH vs [Fe/H] for UMi 
giants from our 
sample (large filled circles). 
The symbol key for the other sources is given on each figure.
Typical uncertainties are shown for one star. 
The line represents the behavior of halo dwarfs from \cite{reddy06}.
\label{figure_cfeh}}
\end{figure}

\begin{figure}
\epsscale{0.9}
\plotone{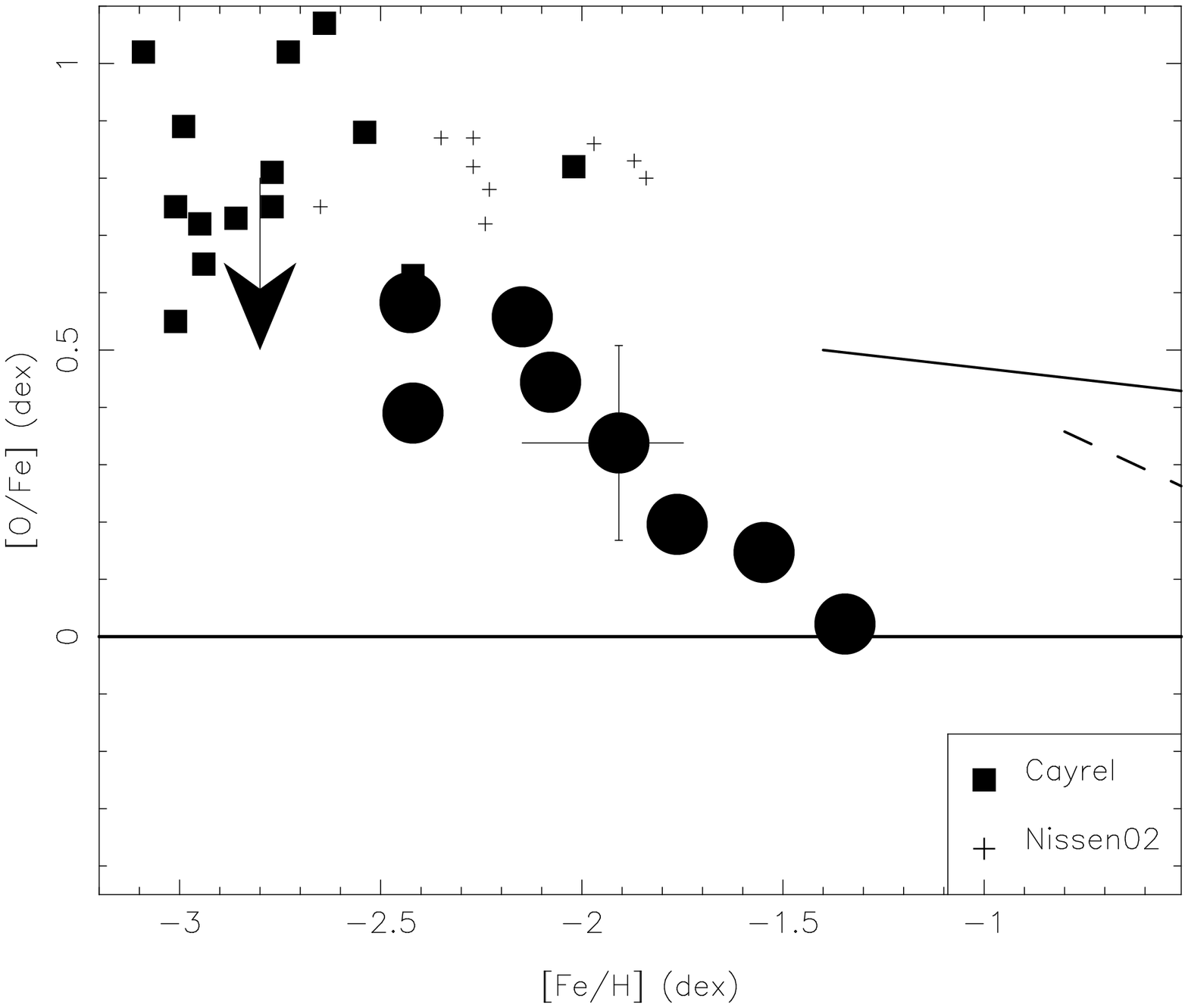}
\caption[]{[O/Fe] vs [Fe/H].  UMi stars from our 
sample are shown as large filled circles.
Typical uncertainties are shown for one star.
The small crosses are from \cite{nissen02} whose sample includes
main sequence and subgiant stars.  Linear fits to the
thick disk and halo stars (solid line) and thin disk (dashed line)
relations of \cite{ramirez07} are shown.  The arrow indicates
the probable magnitude of 1D to 3D model corrections required for
the \cite{cayrel04} and the UMi [O/Fe] values.
\label{figure_ofeh}}
\end{figure}

\begin{figure}
\epsscale{0.9}
\plotone{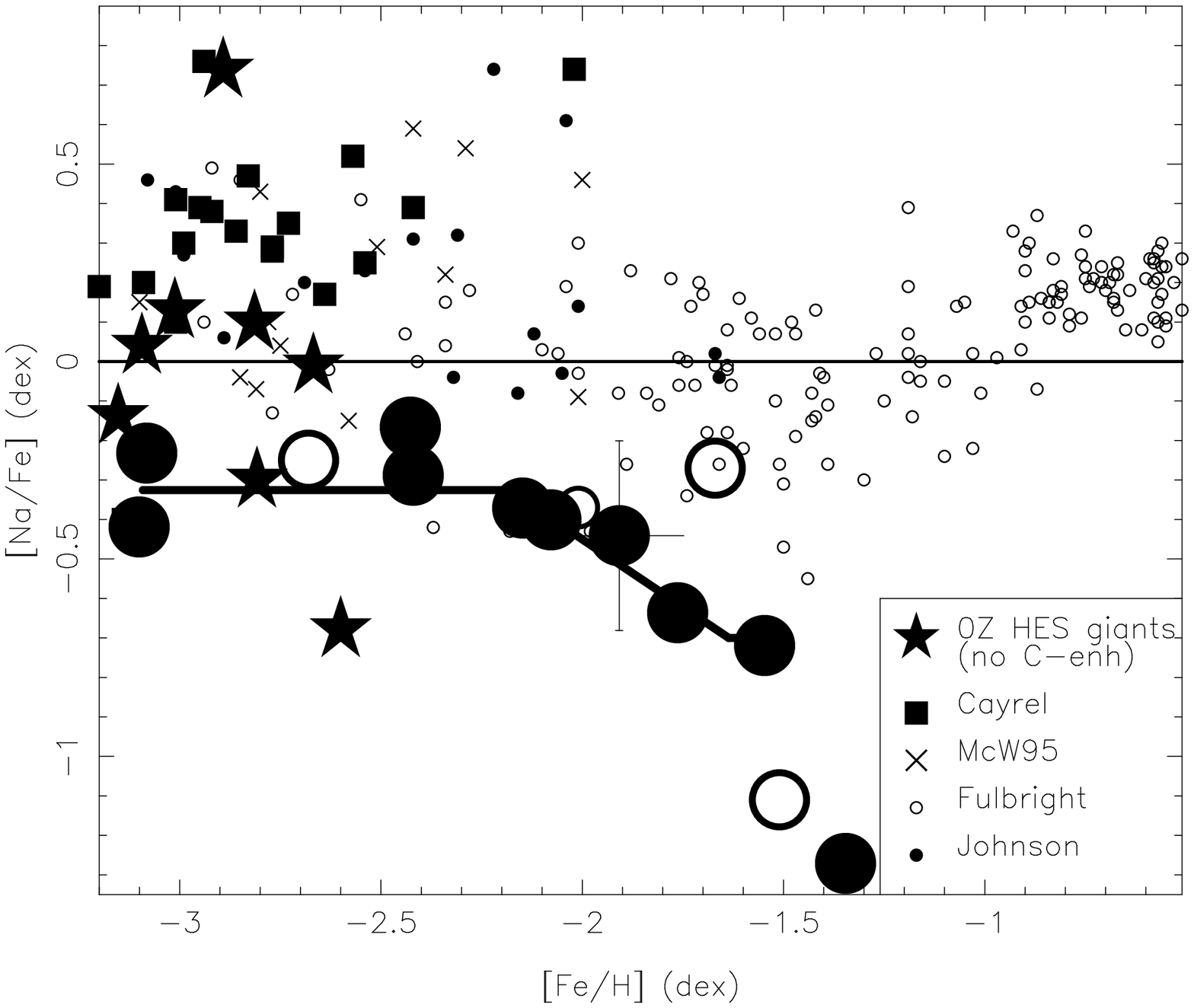}
\caption[]{[Na/Fe] vs [Fe/H] for UMi stars.  
Stars from our HIRES 
sample (large filled circles) are combined 
with those from \cite{subaru} (large open circles). Smaller
open circles denote the somewhat less accurate abundances
from \cite{shetrone2}.  These symbols are used for the rest of
the figures in this paper.
Typical uncertainties are shown for one star.
The symbol key for other sources is shown on the figure.  The thick line
indicates the fit of the toy model described in \S\ref{section_toy_model}
(see also Table~\ref{table_fit})   to the UMi data with COS171 excluded.
\label{figure_nafeh}}
\end{figure}

\begin{figure}
\epsscale{0.9}
%
%
\plotone{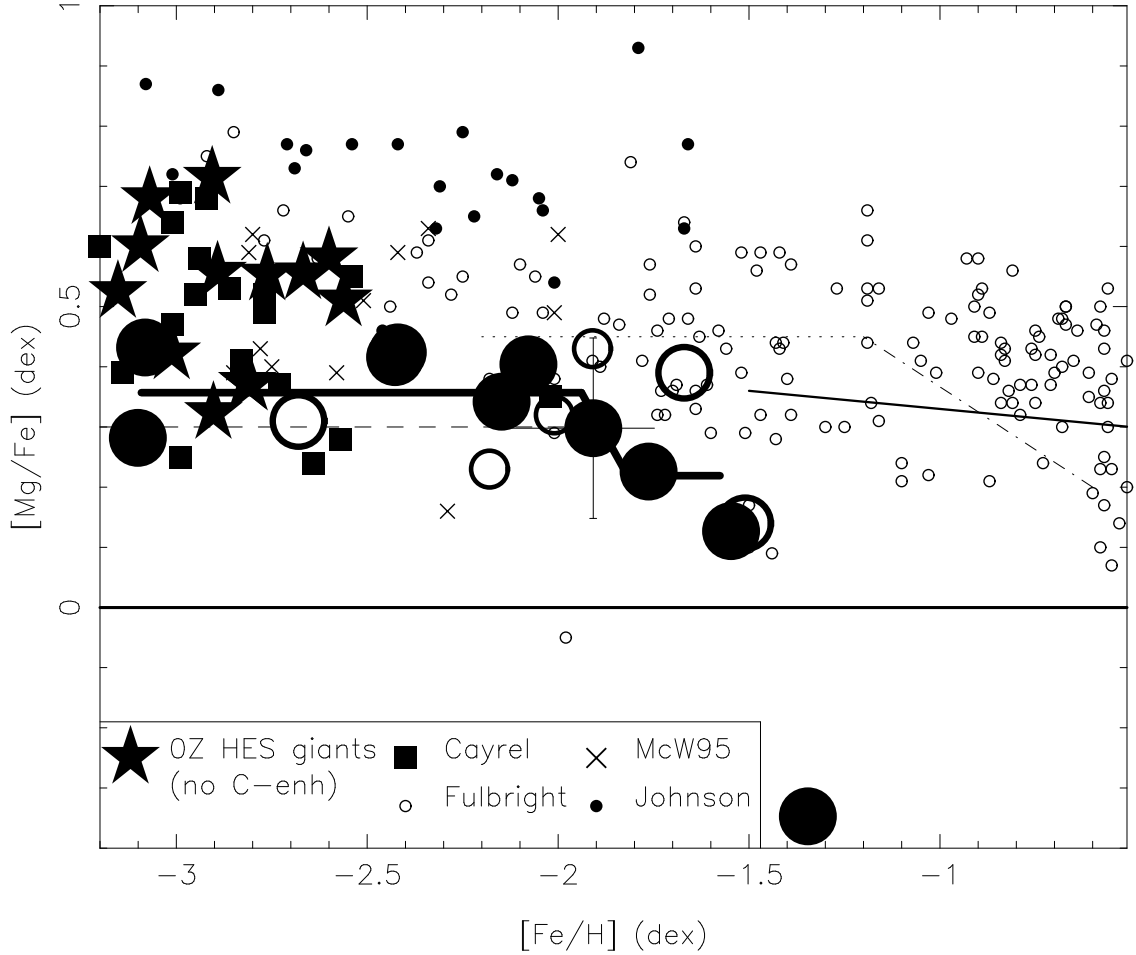}
\caption[]{[Mg/Fe] vs [Fe/H] for the UMi giants.
See Fig.~\ref{figure_nafeh}
for details regarding the symbols for the UMi stars and uncertainties.
The symbol key for sources of data for Galactic halo field
stars is shown on the figure.  Note that 0.15~dex has been added to
the [Mg/Fe] values from \cite{cayrel04}; see the text for details.
The thick line
indicates the fit of the toy model described in \S\ref{section_toy_model}
(see also Table~\ref{table_fit})   to the UMi data with COS171 excluded.
The solid line is the mean relation for thick disk stars
from \cite{reddy06}.  The dotted line is the mean relation for 
inner halo stars from \cite{roederer08}, while his outer halo mean
is shown as the dashed line. 
\label{figure_mgfeh}}
\end{figure}

\begin{figure}
\epsscale{0.9}
\plotone{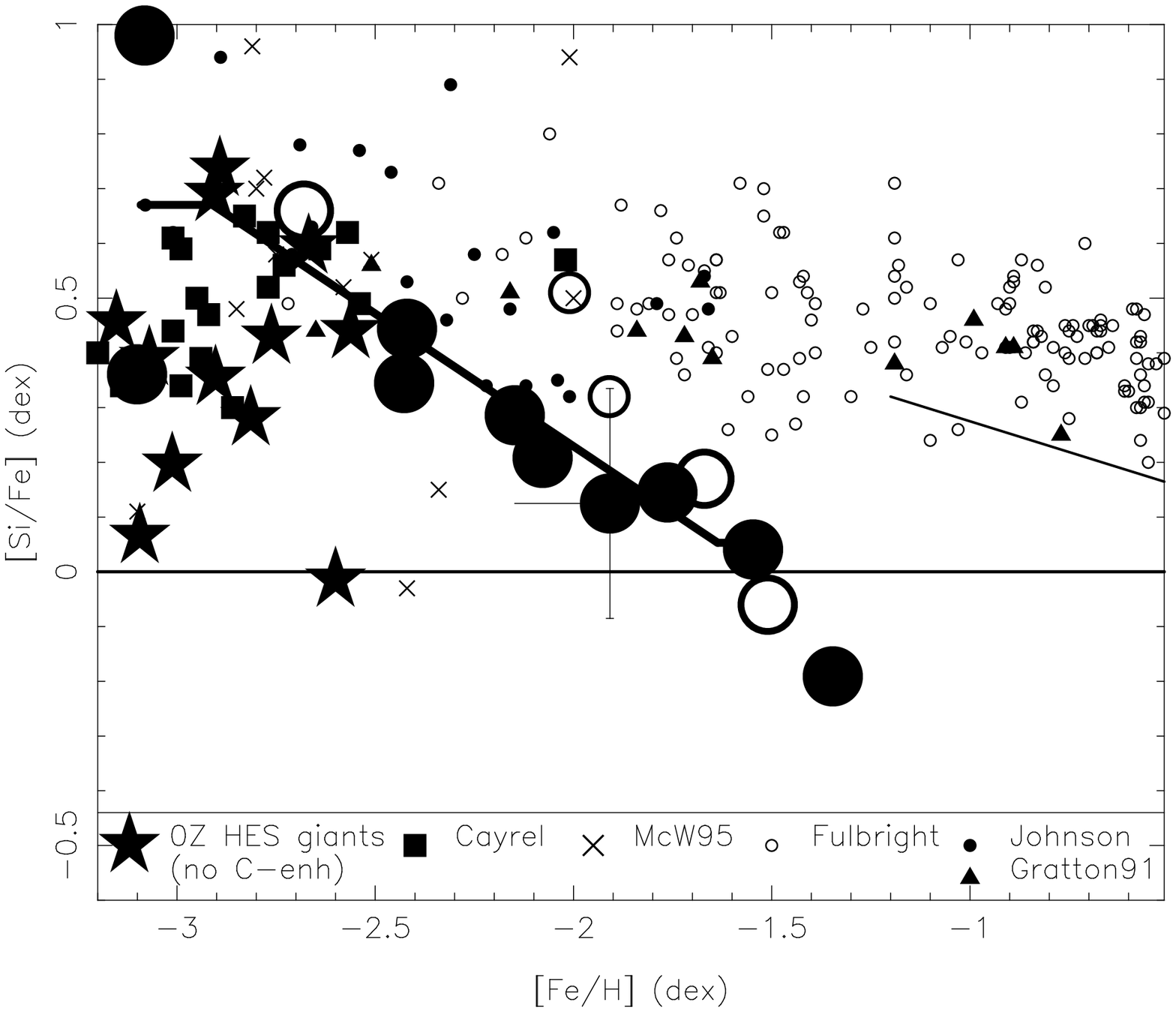}
\caption[]{[Si/Fe] vs [Fe/H] for UMi stars.
See Fig.~\ref{figure_nafeh}
for details regarding the symbols for the UMi stars and uncertainties.
The symbol key for sources of data for Galactic halo field
stars is shown on the figure. 
The thick line
indicates the fit of the toy model described in \S\ref{section_toy_model}
(see also Table~\ref{table_fit})   to the UMi data.  
The solid line is the mean 
relation for thick disk stars from \cite{reddy06}.
\label{figure_sifeh}}
\end{figure}

\begin{figure}
\epsscale{0.9}
\plotone{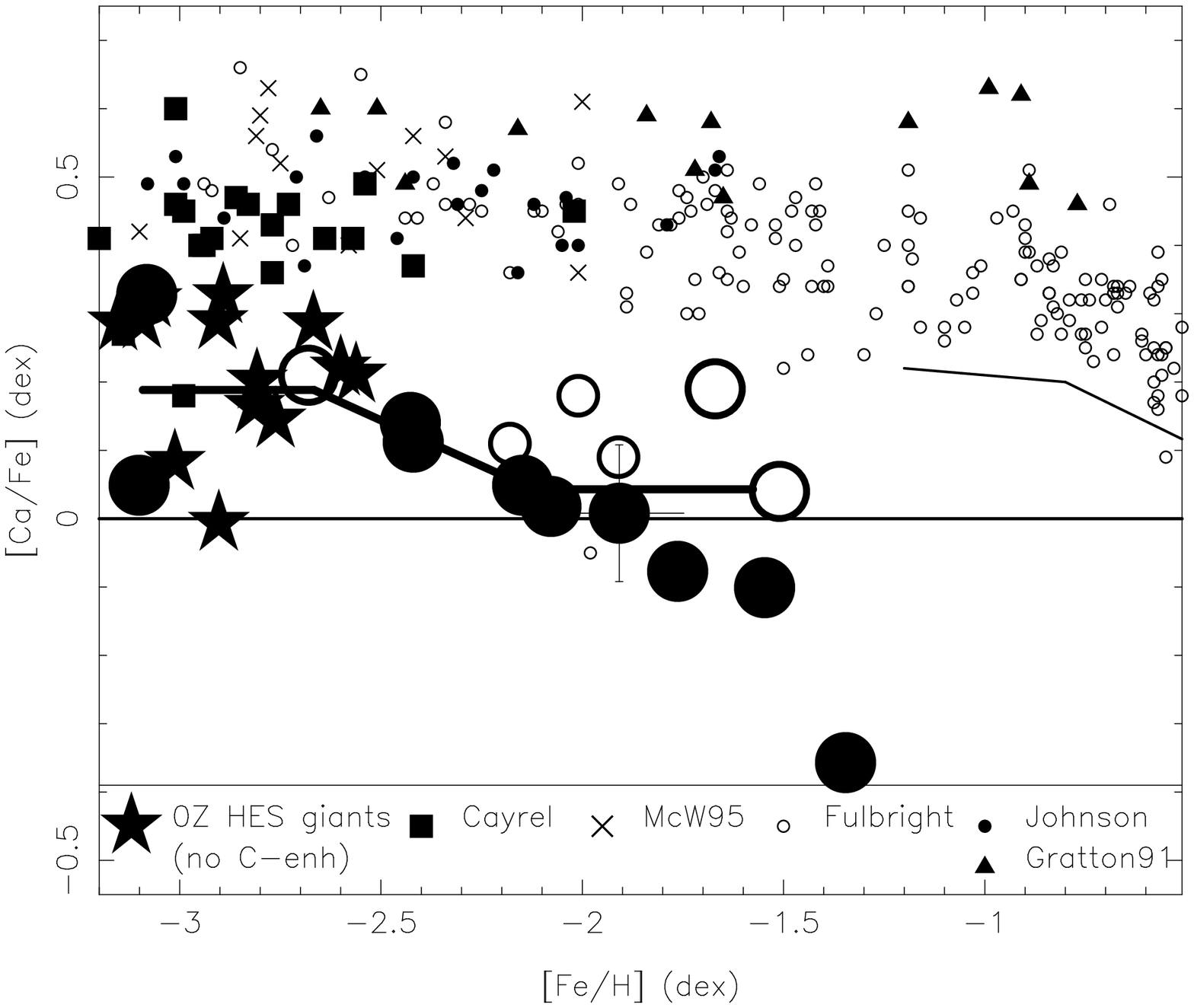}
\caption[]{[Ca/Fe] vs [Fe/H] for UMi stars.
See Fig.~\ref{figure_nafeh}
for details regarding the symbols for the UMi stars and uncertainties.
The symbol key for sources of data for Galactic halo field
stars is shown on the figure. 
The thick line
indicates the fit of the toy model described in \S\ref{section_toy_model}
(see also Table~\ref{table_fit})   to the UMi data  data with COS171 excluded.
The solid line is the mean 
relation for thick disk stars from \cite{reddy06}.
\label{figure_cafeh}}
\end{figure}

\begin{figure}
\epsscale{0.9}
\plotone{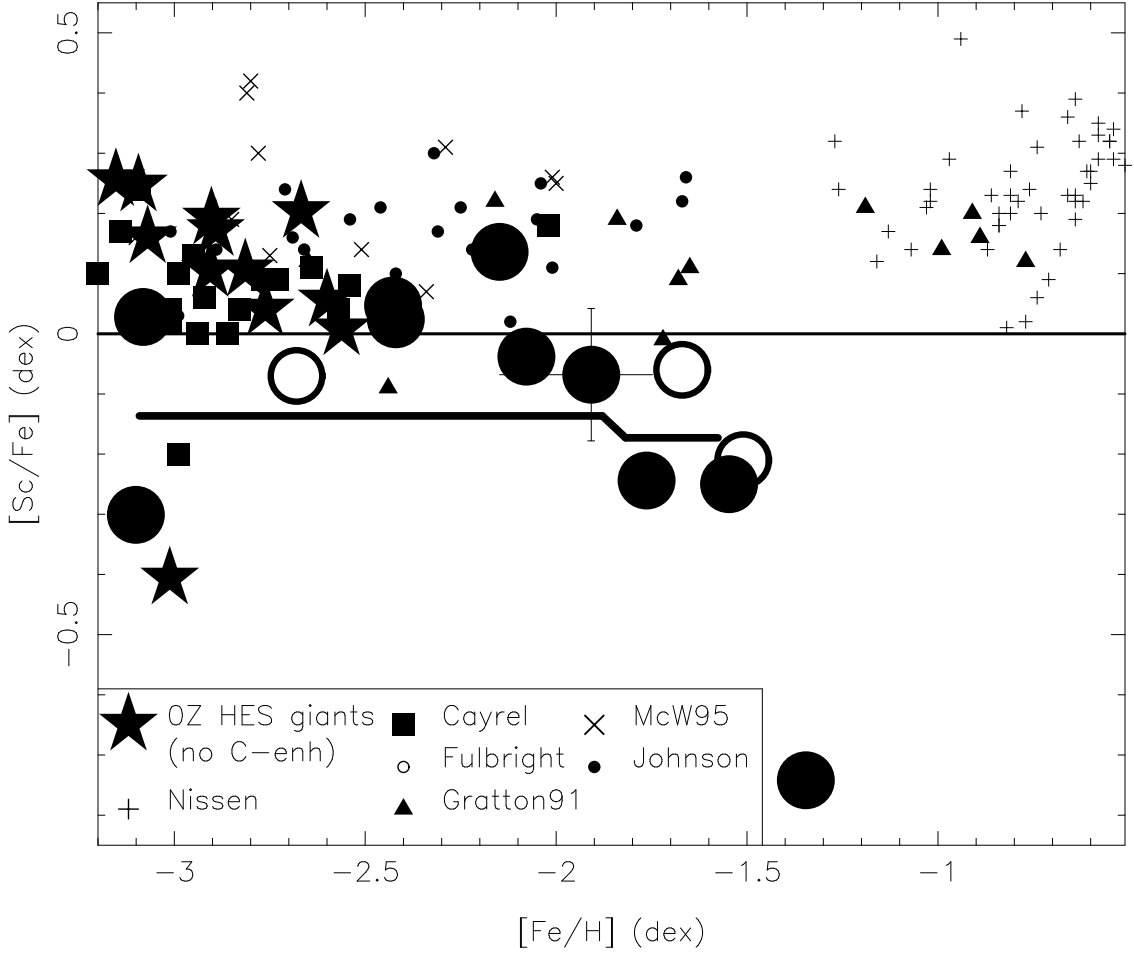}
\caption[]{[Sc/Fe] vs [Fe/H] for UMi giants.  See Fig.~\ref{figure_nafeh}
for details regarding the symbols for the UMi stars and uncertainties.
The thick line
indicates the fit of the toy model described in \S\ref{section_toy_model}
(see also Table~\ref{table_fit})   to the UMi data  data with COS171 excluded.
The symbol key for sources of data for Galactic halo field
stars is shown on the figure.  
\label{figure_scfeh}}
\end{figure}

\begin{figure}
\epsscale{0.9}
\plotone{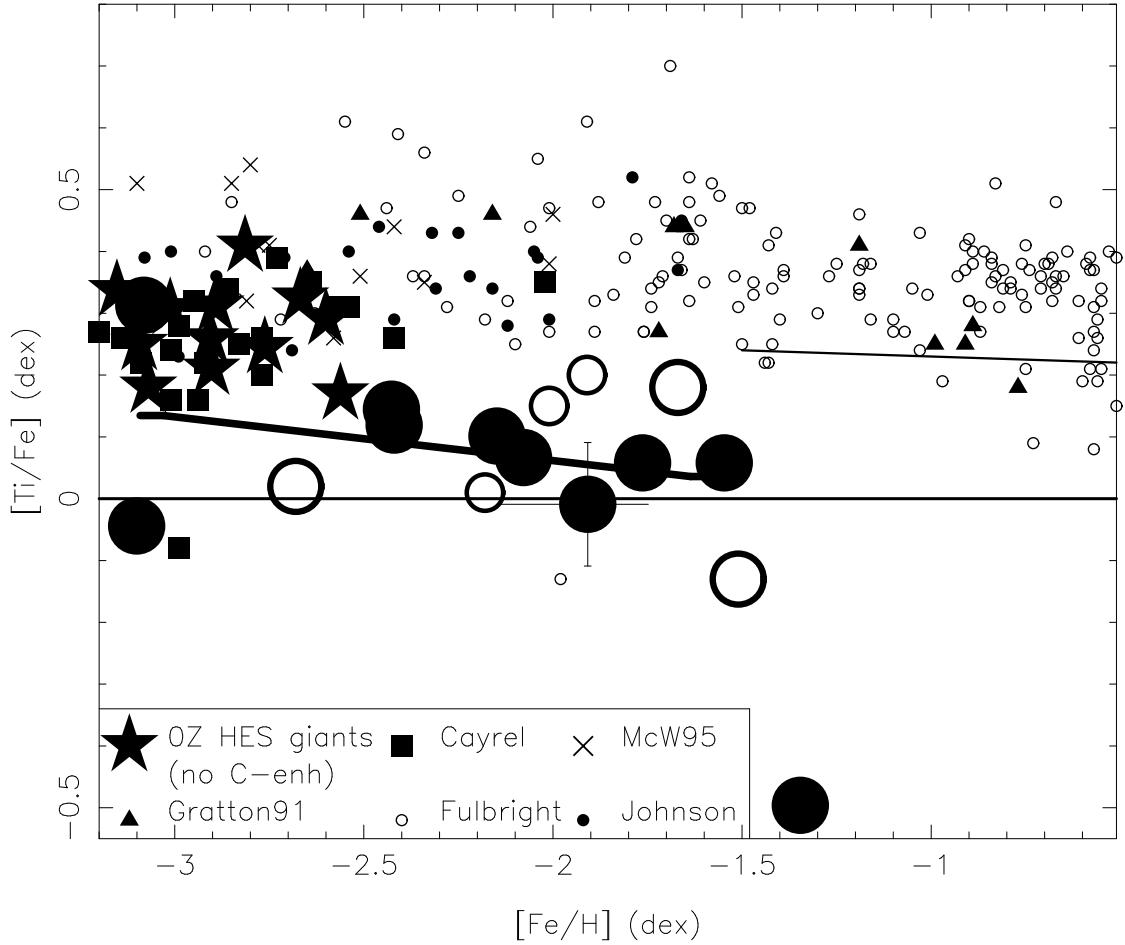}
\caption[]{[Ti/Fe] vs [Fe/H] for UMi stars.
[Ti12/Fe12], which relates ionized Ti to ionized Fe, and neutral
Ti to Fe~I, is shown for our UMi stars.
See Fig.~\ref{figure_nafeh}
for details regarding the symbols for the UMi stars and uncertainties.
The symbol key for sources of data for Galactic halo field
stars is shown on the figure.  The thick line
indicates the fit of the toy model described in \S\ref{section_toy_model}
(see also Table~\ref{table_fit})   to the UMi data with COS171 excluded.
The solid line denotes
the mean relation for the thick disk stars from \cite{reddy06}.
\label{figure_tifeh}}
\end{figure}

\begin{figure}
\epsscale{0.9}
\plotone{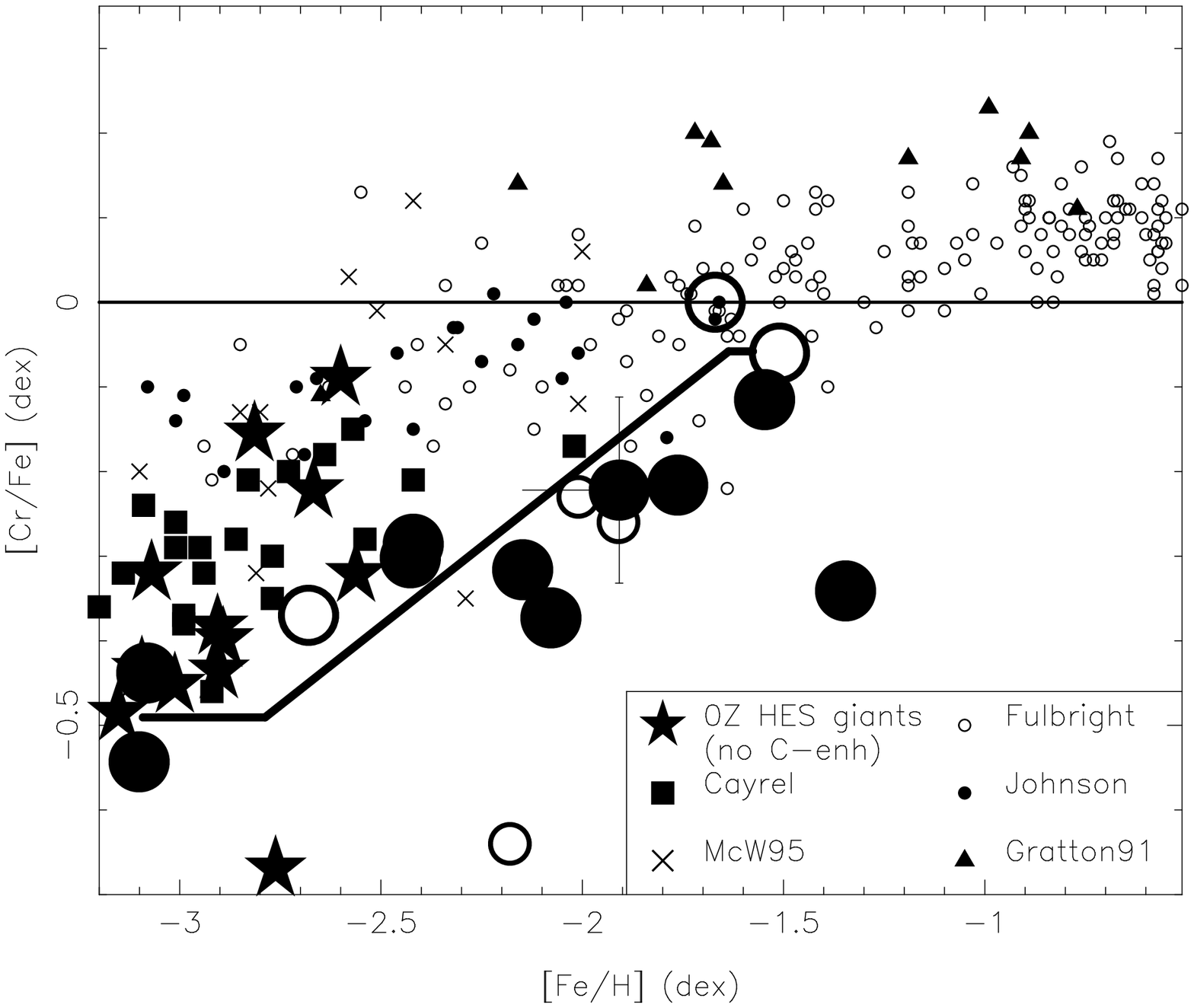}
\caption[]{[Cr/Fe]  vs [Fe/H] for UMi stars.
See Fig.~\ref{figure_nafeh}
for details regarding the symbols for the UMi stars and uncertainties.
The symbol key for sources of data for Galactic halo field
stars is shown on the figure.  The thick line
indicates the fit of the toy model described in \S\ref{section_toy_model}
(see also Table~\ref{table_fit})   to the UMi data with COS171 excluded.
\label{figure_crfeh}}
\end{figure}

\begin{figure}
\epsscale{0.9}
\plotone{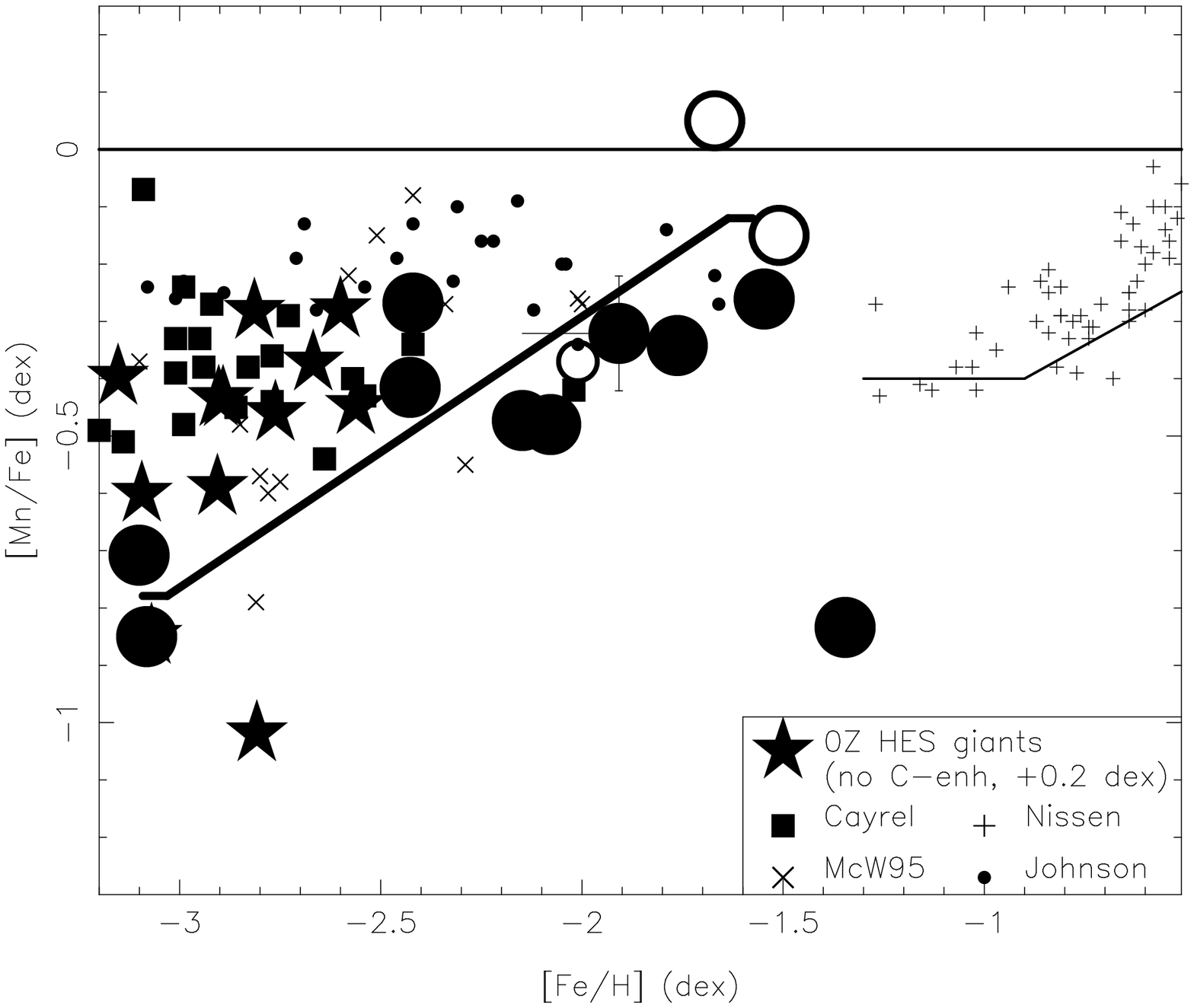}
\caption[]{[Mn/Fe] vs. [Fe/H] for UMi giants.
See Fig.~\ref{figure_nafeh}
for details regarding the symbols for the UMi stars and uncertainties.
The symbol key for sources of data for Galactic halo field
stars is shown on the figure.  The thick line
indicates the fit of the toy model described in \S\ref{section_toy_model}
(see also Table~\ref{table_fit})   to the UMi data with COS171 excluded.
The solid line denotes
the mean relation for the thick disk stars from \cite{reddy06}.
\label{figure_mnfeh}}
\end{figure}

\begin{figure}
\epsscale{0.9}
\plotone{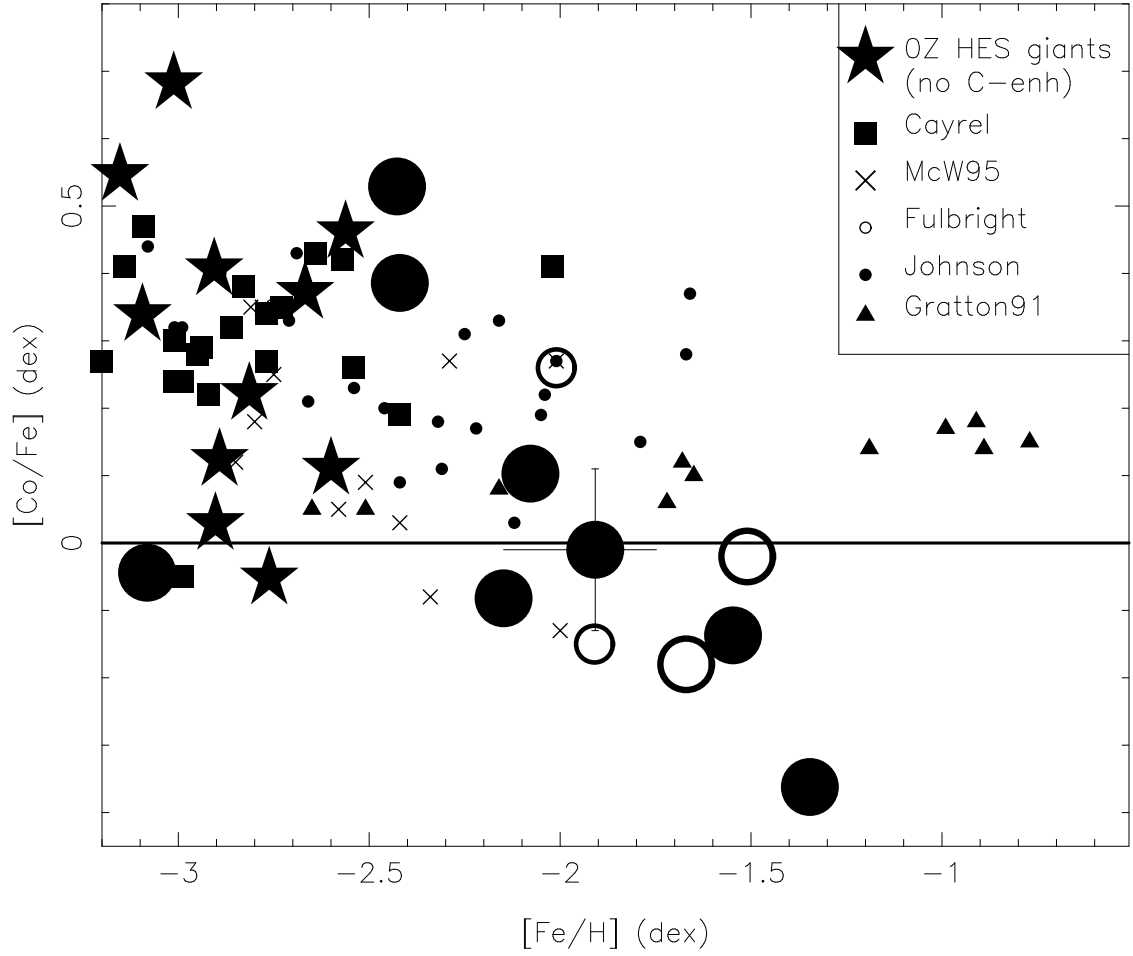}
\caption[]{[Co/Fe]  vs [Fe/H] for UMi stars.
See Fig.~\ref{figure_nafeh}
for details regarding the symbols for the UMi stars and uncertainties.
The symbol key for sources of data for Galactic halo field
stars is shown on the figure.
\label{figure_cofeh}}
\end{figure}

\begin{figure}
\epsscale{0.9}
\plotone{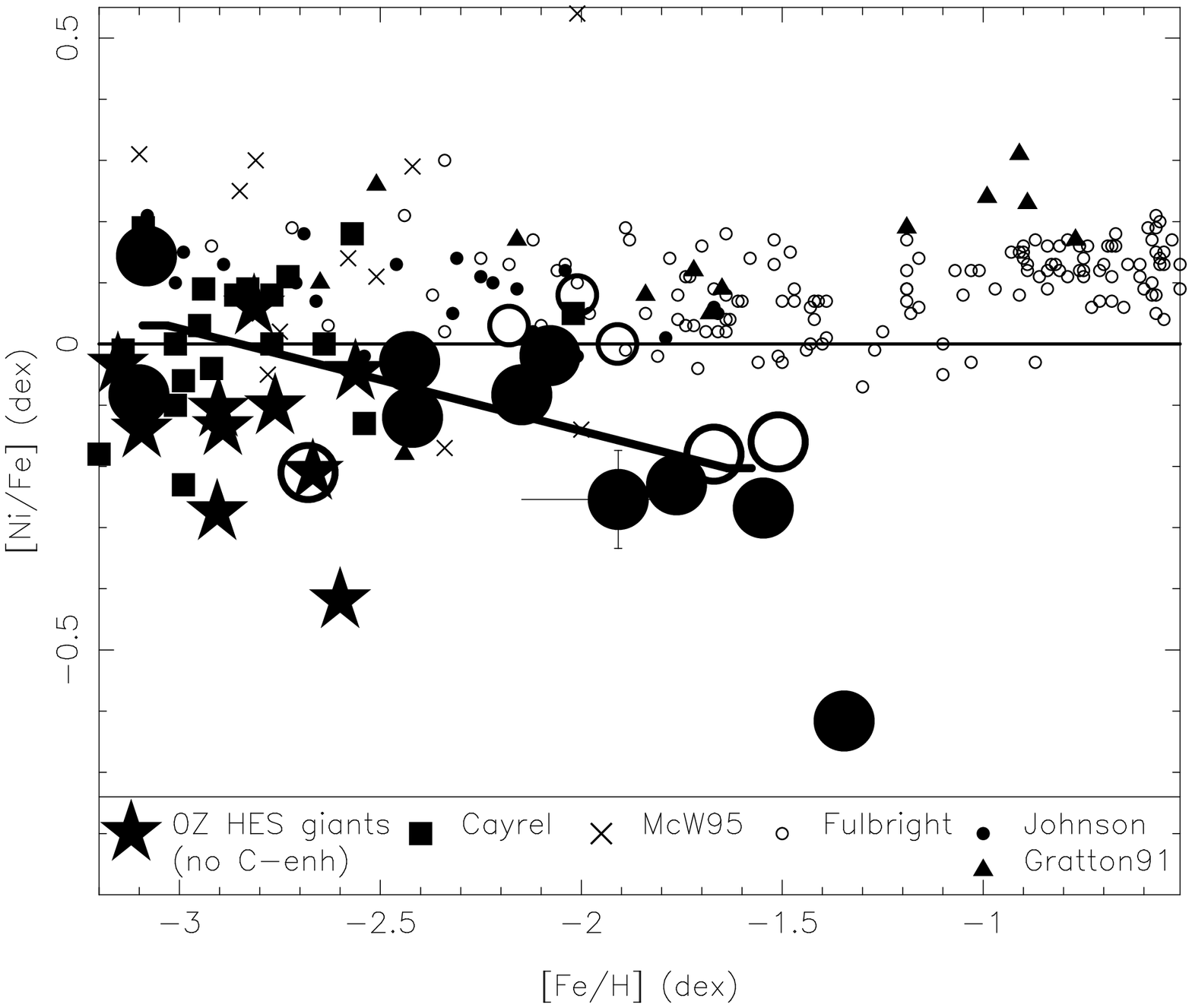}
\caption[]{[Ni/Fe]  vs [Fe/H] for UMi stars.
See Fig.~\ref{figure_nafeh}
for details regarding the symbols for the UMi stars and uncertainties.
The symbol key for sources of data for Galactic halo field
stars is shown on the figure.  The thick line
indicates the fit of the toy model described in \S\ref{section_toy_model}
(see also Table~\ref{table_fit})  to the UMi data with COS171 excluded.
\label{figure_nifeh}}
\end{figure}

\begin{figure}
\epsscale{0.9}
\plotone{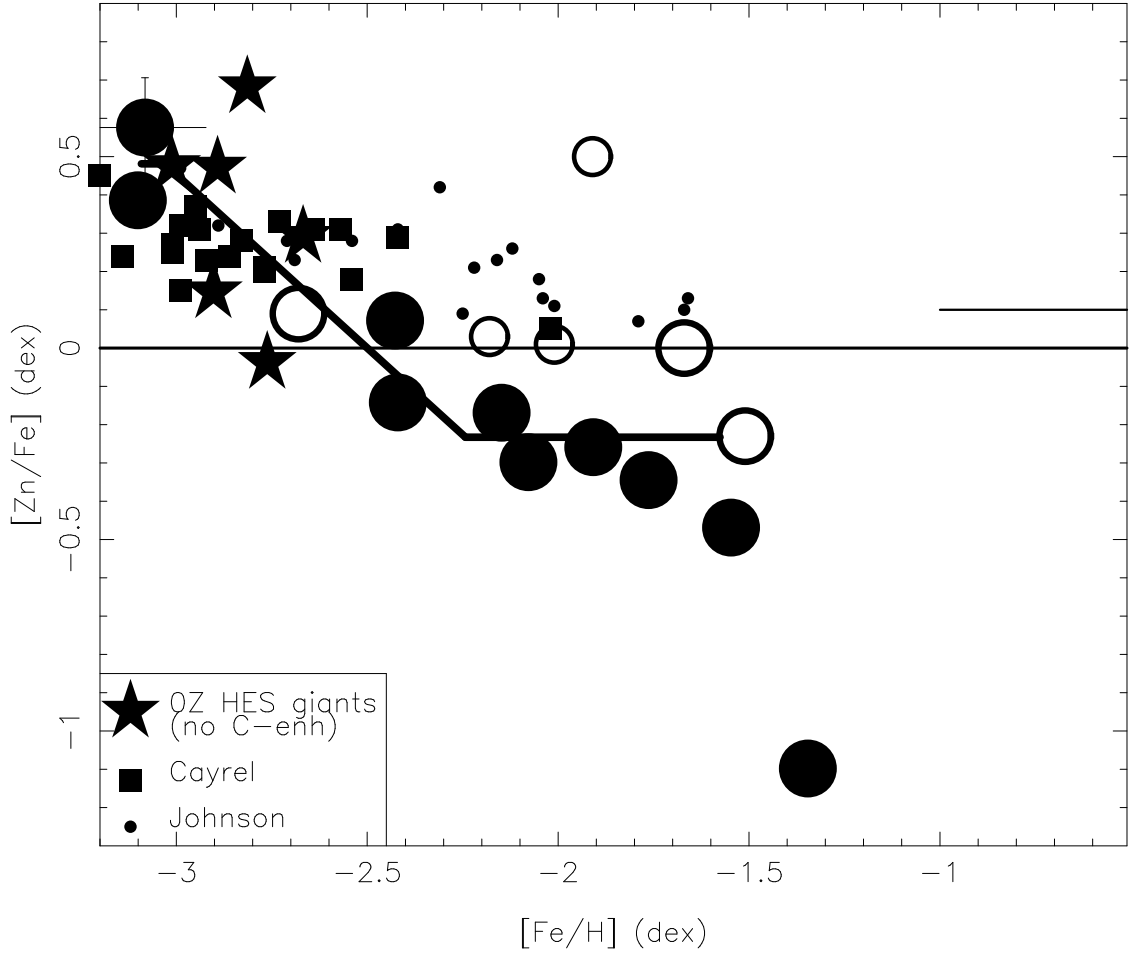}
\caption[]{[Zn/Fe]  vs [Fe/H] for UMi stars.
See Fig.~\ref{figure_nafeh}
for details regarding the symbols for the UMi stars and uncertainties.
The symbol key for sources of data for Galactic halo field
stars is shown on the figure. The thick line
indicates the fit of the toy model described in \S\ref{section_toy_model}
(see also Table~\ref{table_fit})   to the UMi data with COS171 excluded.
  The behavior
of this abundance ratio in thick disk dwarfs from \cite{reddy06}
is indicated as a solid line.
\label{figure_znfeh}}
\end{figure}

\begin{figure}
\epsscale{0.9}
\plotone{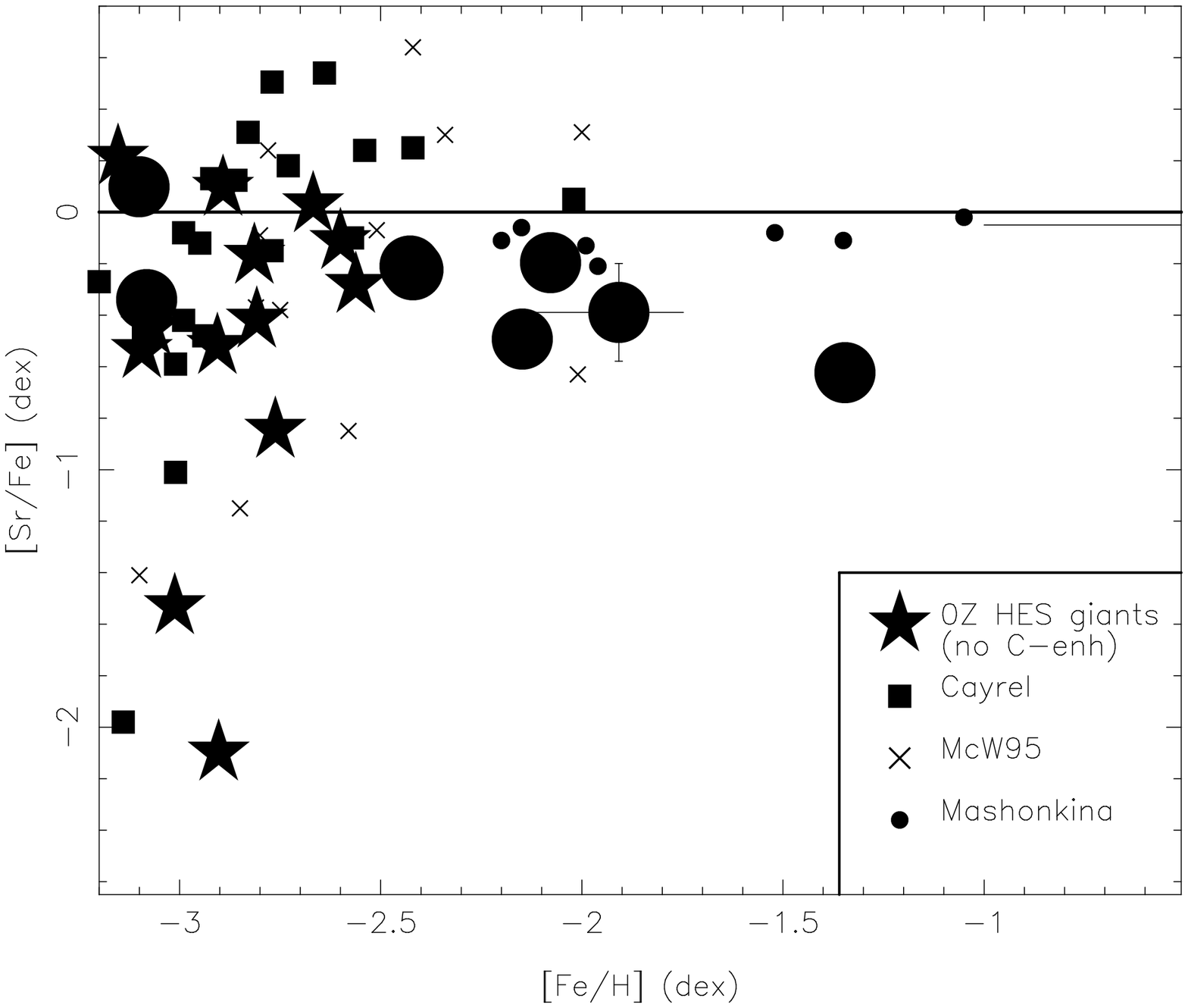}
\caption[]{[Sr/Fe]  vs [Fe/H] for UMi stars.
See Fig.~\ref{figure_nafeh}
for details regarding the symbols for the UMi stars and uncertainties.
The First Stars data is from \cite{francois}.
The symbol key for sources of data for Galactic halo field
stars is shown on the figure. 
The thick line
indicates the fit of the toy model described in \S\ref{section_toy_model}
(see also Table~\ref{table_fit})  to the UMi data.
The behavior
of this abundance ratio in thick disk dwarfs from 
\cite{halo_sr} is shown as the solid line.
\label{figure_srfeh}}
\end{figure}

\begin{figure}
\epsscale{0.9}
\plotone{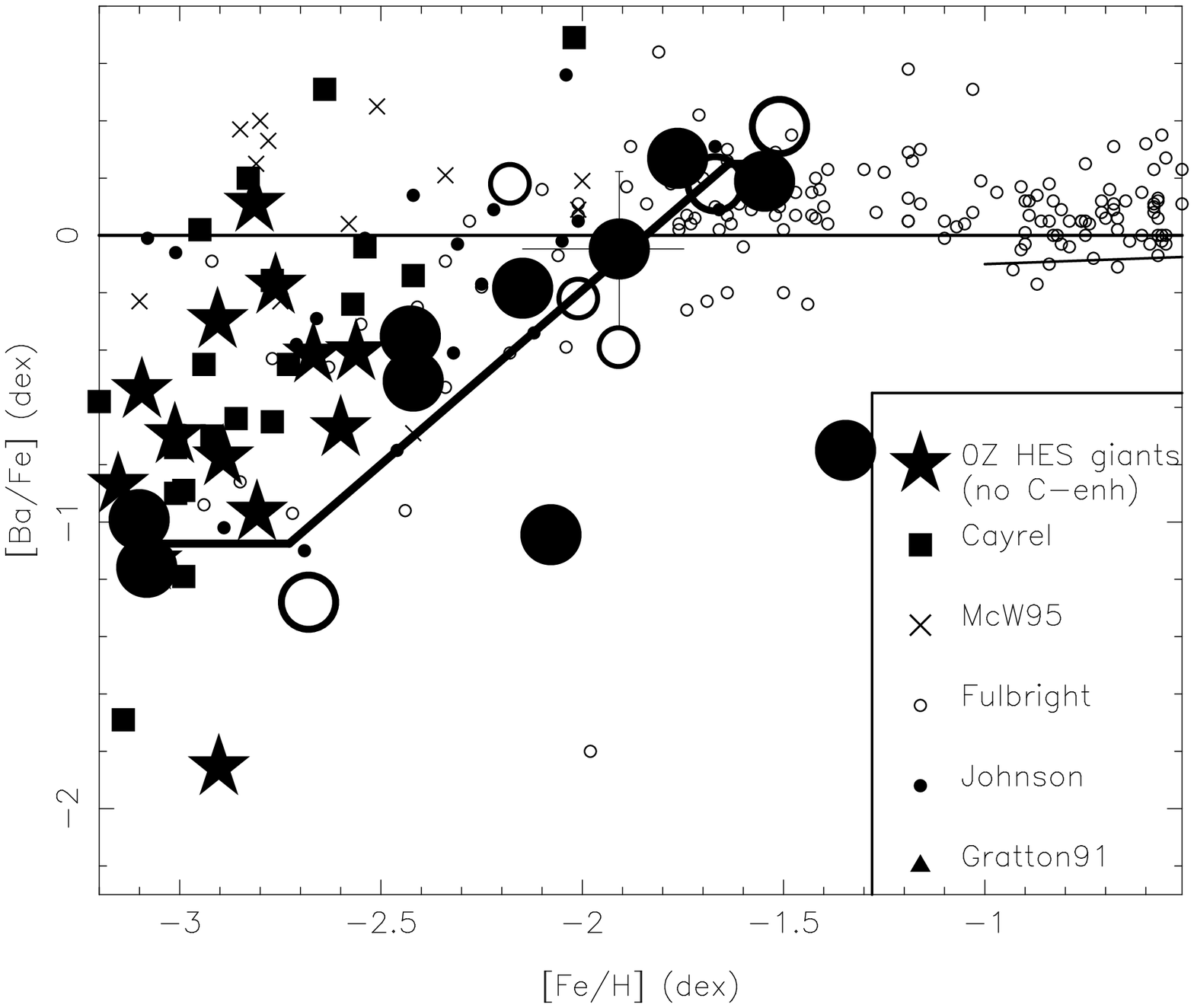}
\caption[]{[Ba/Fe]  vs [Fe/H] for UMi stars.
See Fig.~\ref{figure_nafeh}
for details regarding the symbols for the UMi stars and uncertainties.
The First Stars data is from \cite{francois}.
The symbol key for sources of data for Galactic halo field
stars is shown on the figure.  The thick line
indicates the fit of the toy model described in \S\ref{section_toy_model}
(see also Table~\ref{table_fit})  to the UMi data with COS171 excluded.
\label{figure_bafeh}}
\end{figure}

We proceed  by examining a series of plots 
(Figs.~\ref{figure_cfeh} to \ref{figure_bafeh}) in which we show
the UMi sample, both our 10 stars (indicated by large
filled circles) and the 6 observed
previously from  \cite{shetrone2} (denoted by small
open circles, and less accurate than subsequent UMi studies), with
the more accurate \cite{subaru} abundances indicated by
large open circles. 
These figures
also display current results for Galactic halo stars, but see also
the seminal early review of \cite{mcwilliam97}.
The main halo survey included is
the 0Z project led by J.~Cohen
to datamine the Hamburg/ESO Survey for extremely metal-poor
stars in the Galactic halo.  Many of the most metal-poor candidates from this
work have been observed with HIRES at the Keck Observatory
and analyzed in a manner very similar to the present study
as described in \cite{cohen04} and \cite{cohen08},
with the difference that most of the spectra for the 0Z project
were taken further towards the blue than those of the dSph stars,
a move necessary because of the low density of lines in the red
in spectra of such low
metallicity stars.   Only the giants in the metal-rich end of
the 0Z project database, much of which is not yet published
(J.~Cohen, N.~Christlieb et al, in preparation),
is shown in these figures. 
A number of other halo field star surveys,
the most important of which at the lower metallicities probed here
is the First Stars Project  \citep{cayrel04}, are shown, 
including
those stars from \cite{mcwilliam95a}
not re-observed by \cite{cayrel04}. 

It should be noted that these Galactic halo
field star surveys are dominated by inner halo
stars with $R_{GC} < 20$~kpc adopting  the 
inner/outer halo boundary set by \cite{carollo}.  If one 
redefines this boundary
to lie at a somewhat smaller $R_{GC}$, then many of the 0Z giants
are in the outer halo.      
Much smaller samples of probable outer halo dwarfs in the local neighborhood
have been isolated from their kinematics, and their chemical inventory analyzed in
detail in several previous studies, in particular
by \cite{nissen97,nissen10} and by \cite{stephens99}.  \cite{roederer08}
has compiled a sample of halo stars with parallaxes to isolate
outer halo stars.   These studies collectively find 
a small deficit in [Mg/Fe] in outer halo stars as compared to inner
halo ones shown by the dotted and dashed lines in Fig.~\ref{figure_mgfeh},
accompanied by slightly subsolar [Na/Fe] and [Ni/Fe].

We will see that the differences
in the chemical inventory
between Galactic halo field stars and the UMi sample,
which may be a function of Fe-metallicity,  
are small, not larger than $\sim$0.3~dex
in most cases.  This means that some care is required
to ensure that all the abundances from the various sources
are homogeneous.  While we have not done a full check of this,
we have taken a few steps the first of which is
to adjust each survey to our set of Solar abundances,
particularly to our adopted value of [Fe/H], whenever
possible. Specific
cases where there are clear problems related to issues
of homogeneity between the various analyses are noted individually below.

Overall the abundance relations we find here
for the UMi augmented sample are 
more clearly defined with less scatter than we found earlier
for Draco.  In part this is a consequence of the (small) difference
in distance, with UMi being somewhat closer, hence having somewhat
brighter stars near the RGB tip, resulting in better spectra.  But we
wonder if part of this is also a result of the more extended epoch of
star formation in Draco than in UMi, resulting in a more complex
chemical evolution with stronger spatial variations within the Draco dSph.

The trend of [C/Fe] versus [Fe/H] is shown in Fig.~\ref{figure_cfeh}, based
for the majority of these stars on the strength of the G band of CH.
The solid lines represent the mean behavior of thick disk dwarfs 
from the survey by \cite{reddy06}. 
The C abundance in luminous giants is lowered substantially
from an initial [C/Fe] $\approx$ 0.0~dex 
due to intrinsic nucleosynthesis (the CN cycle of H burning)
followed by dredge-up to the stellar surface of processed material
within which C has burned to N
\citep[see e.g.][]{cohen_m15}.  The C abundances in most of the
luminous UMi giants studied here are abnormally low, presumably
due to mixing;  their  initial
C abundances  cannot be determined.
Note that in Fig.~\ref{figure_cfeh} and those that follow the asymmetric
uncertainties we adopted in \S\ref{section_param}
are shown for [Fe/H], but not for abundance ratios
[X/Fe]; for the latter the larger uncertainty is plotted.

It is quite difficult to measure  O abundances in metal-poor giants. 
The set of features that can be used
is very limited and each has problems.
This has resulted in considerable
controversy about O abundances in metal-poor stars in recent years,
see e.g. the discussion in \cite{melendez06}.
The forbidden OI lines at 6300 and 6363~\AA\ line are very weak, and
the 7770~\AA\ triplet, which has substantial non-LTE effects, 
is not detectable.   [O/Fe] ratios for the UMi giants
and for a compilation of surveys in the literature are shown
in Fig.~\ref{figure_ofeh}.

The arrow in Fig.~\ref{figure_ofeh} indicates the probable
correction for 1D to 3D effects required for luminous
giants given by \cite{cayrel04}, which has not been implemented,
but which would bring the plateau in [O/Fe] down to a mean
level of $\sim$+0.5~dex. 
The lines are linear fits from \cite{ramirez07} to their
samples of
thick disk and halo dwarfs (solid line) and to thin
disk dwarfs (dashed line).  They use only 
the 7770~\AA\ triplet, with appropriate non-LTE corrections;
these lines become detectable in dwarf stars but are considerably
weaker in giants.
The net result is that the
UMi giants appear low in [O/Fe] when compared to 
samples of field halo giants which rely on the same 6363~\AA\
forbidden line.

Fig.~\ref{figure_nafeh} shows that
the
UMi giants clearly have [Na/Fe] somewhat lower than the
Galactic halo field stars over the entire metallicity
range spanned within UMi, a trend seen at intermediate
metallicities for outer halo local dwarfs by
\cite{nissen10}.  There is a very large range in
[Na/Fe] among the highest Fe-metallicity stars in UMi.
\cite{spite} found that Na/Fe ratios vary by a factor of $\sim$5
from star to star among very metal-poor luminous RGB stars, 
which they interpret as  as a result
of deep mixing.
The figure suggests
that there is a separation
of $\sim$0.2~dex for [Na/Fe]  at a
fixed [Fe/H] between the two large surveys of very metal-poor
halo field stars, i.e. the First Stars
Survey led by R.~Cayrel and the 0Z Survey led by J.~Cohen. 
\cite{andrievsky07} have demonstrated that non-LTE effects
in [Na/Fe] based solely on the NaD lines are substantial and
depend on the luminosity
and \teff\ of the star.  Hence part of the origin 
of this difference for [Na/Fe] may arise from a difference in
mean sample luminosity between these two surveys of halo field
stars; see C09 for additional discussion.

Fig.~\ref{figure_mgfeh} shows the
important hydrostatic $\alpha$-element Mg, another element
with only a few accessible features in our UMi spectra.
The published values from \cite{cayrel04} for
the First Stars project have been increased by 0.15~dex 
following \cite{bonifacio09}.
We find that  [Mg/Fe] is constant to within the uncertainties
at the super-solar value of
$\sim$0.35~dex, consistent
with that typical of outer halo Galactic giants
found by \cite{roederer08}, at all Fe-metallicities 
among the UMi giants.  The highest
[Fe/H] UMi giant, COS171, has [Mg/Fe] 0.5~dex lower than the
three other stars of similar [Fe/H].  It is a low
outlier for this and for many other species and is discussed in 
\S\ref{section_outlier}.

The  behavior of the explosive $\alpha$-element Si is
shown in Fig.~\ref{figure_sifeh} with the mean relation for thick
disk stars from \cite{reddy06} indicated.  The figure shows
good agreement  between the 0Z and First Stars Project
abundance ratios for this element. The 
lowest Fe-metallicity UMi giants show [Si/Fe] consistent with
that of Galactic halo stars, but this ratio falls steadily
with increasing [Fe/H] in UMi, while it remains constant
among the halo stars.   The Solar ratio of [Si/Fe] is
reached at [Fe/H] $\sim -1.6$~dex, far more metal-poor than
is typical of Galactic populations.

The explosive $\alpha$-element Ca also has problems with inconsistencies
between the two large surveys of very metal-poor halo stars,
the First Stars Project and the 0Z Project; this issue is discussed in C09.
No detectable difference between the inner and outer halo
was found by \cite{roederer08} or \cite{ishigaki10}, so the
mean distance of the halo sample is not relevant.
Ignoring the low outlier
COS171, [Ca/Fe] is +0.1$\pm0.1$~dex for all the UMi giants.
If the 0Z measurements of [Ca/Fe] are adopted, then
the luminous UMi giants have [Ca/Fe] comparable to, or
only slightly lower than, those
of Galactic halo stars over the full
range of [Fe/H] found in UMi.

Figs.~\ref{figure_scfeh} and \ref{figure_tifeh}
show the behavior for Sc and  the explosive $\alpha$-element Ti
respectively.
The mean relation for thick
disk stars from \cite{reddy06} is shown for the latter.
In both cases there is good agreement between the
abundance ratios deduced by
the 0Z Project and the First Stars Project.
[Sc/Fe] is slightly sub-solar 
and below the Galactic halo field giants over the full range of 
[Fe/H].
For the explosive $\alpha$-element
Ti  the
metal-rich UMi stars are slightly above the  solar value,
but fall  below the
halo field. [Ti/Fe] at the extremely metal-poor end of
the UMi sample may be closer to the halo field, but the value
is uncertain there.

In UMi it is the explosive $\alpha$-element Si which shows the
strongest divergence from the Galactic halo field as a function of
increasing [Fe/H].  The hydrostatic element [Mg/Fe] behaves fairly
close to the outer halo trends of \cite{roederer08}.
The small range in [Mg/Fe] seen among the UMi giants is in contrast
to Draco (see C09), where there is a stronger decrease as [Fe/H] increases.
The latter might be expected for a more extended epoch of star formation
since Mg, unlike Ca or Si, is
produced only in SNII, while Ca and Si are produced
in both SNII and SNIa \citep{woosley}.  However, given the
postulated very short duration of star formation within UMi, the
SNIa never had time to contribute for any element in this dSph,
while in Draco,  star formation lasted  long enough
for some SNIa contribution.

There are several elements which probably have metallicity dependent
yields,  related to the value of the neutron excess,
such that at low metallicity, the yield is reduced,
as appears to be the case for the UMi stars.
This includes
Na, Sc, Mn \citep[see e.g.][]{cescutti08}, Ni, and Zn;
their production is discussed in
\cite{arnett} \citep[see also][]{clayton03} for Na, 
\cite{woosley} and \cite{limongi03} for Sc, 
\cite{woosley} 
and \cite{ohkubo} for Ni, and \cite{timmes} for Zn.
Stronger odd-even effects are found for lower metallicity
and, in the case of Sc, for lower mass progenitors \citep{limongi03}.
Thus a relative absence of the higher mass SNII
with $M > 35M$\subsun\ might
give rise to the low [Sc/Fe] in the UMi and in the Draco sample. 

[Cr/Fe] (Fig.~\ref{figure_crfeh}) and [Mn/Fe] (Fig.~\ref{figure_mnfeh})
for the UMi giants overlaps the lower edge of the
 distribution for Galactic
halo field stars.  
Both of these abundance ratios
decline rapidly 
as [Fe/H] decreases in Galactic halo field stars.  A known problem discussed in
\cite{cohen04} requires that the Mn abundance
derived from the 4030 resonance triplet lines be
increased by 0.2~dex.
These are the strongest Mn~I lines in the optical and the only
ones accessible for extremely metal-poor stars.  The offset
has been applied to the two most metal-poor UMi giants, where these
were the only Mn features detected. 
\cite{mn_nolte} suggest
that the non-LTE corrections for Mn in very metal poor giants 
are large and positive, and will
flatten the [Mn/Fe] ratio to a constant value of about $ -0.1$~dex 
for [Fe/H] $< -1.5$~dex.  Since all the stars used here are luminous
giants, the non-LTE effects will presumably be of comparable size
for every star of a fixed [Fe/H], and hence will not significantly affect
statements regarding relative differences between the UMi giants
and the Galactic halo giants.

Fig.~\ref{figure_cofeh} displays the [Co/Fe] ratios which
for Galactic halo stars
rise rapidly from near the Solar ratio as [Fe/H] decreases below $-2$~dex.
Co was only detected in one of the two EMP stars in our UMi sample,
but at higher [Fe/H], it is slightly subsolar, perhaps somewhat lower than the
halo stars.
However, there is only one Co~I line with equivalent width
exceeding 20~m\AA\
in most of these stars, which is
at 4121~\AA, uncomfortably far in the blue.  Given the paucity of suitable lines, any 
conclusion regarding the behavior of [Co/Fe] in UMi is 
still uncertain. The large positive non-LTE corrections suggested by
\cite{co_nolte} further complicate the situation.

The nickel abundance relative to Fe (Fig.~\ref{figure_nifeh}) 
appears to fall below that of the
halo field (which has [Ni/Fe] at the Solar ratio over the
entire range of Fe-metallicity) among the higher metallicity
UMi stars. \cite{nissen10} suggest that in the outer
halo [Ni/Fe] is slightly subsolar. The [Ni/Fe] ratios
for the lowest metallicity UMi stars overlap
in Fig.~\ref{figure_nifeh} those of  Galactic field halo stars.

The Galactic halo field samples from the 0Z Project and the
First Stars Project overlap well for the abundance ratio [Zn/Fe].
Among field halo stars, [Zn/Fe] is close to the Solar ratio
but rises rapidly below [Fe/H] $\sim -2$~dex, as shown
most recently for halo dwarfs by \cite{nissen08}.  In the UMi giants,
[Zn/Fe] behaves similarly to the Ni abundance ratio
for intermediate metallicities;  the UMi stars
fall below those in the Galactic halo and below the Solar ratio
in this regime of [Fe/H].  
At the lowest
metallicities, [Zn/Fe] for the UMi giants appears to
rise above the Solar value.  

The Galactic halo field samples from the 0Z Project and the
First Stars Project \citep[data for Sr and Ba is from][]{francois} 
overlap well for the abundance ratio [Sr/Fe]
versus Fe-metallicity shown in Fig.~\ref{figure_srfeh}. 
The Sr~II lines used are
the resonance lines at 4077 and 4215~\AA; they are
uncomfortably far
in the blue for the UMi spectra, where the S/N  is rather low.
The limited detections of these lines for the UMi giants,
including those with [Fe/H] $< -3$~dex, suggests
that [Sr/Fe] is approximately constant at $-0.1$~dex.
This is in good agreement with the behavior of the bulk of the halo field
star samples.
Fig.~\ref{figure_bafeh} shows the abundance ratios [Ba/Fe],
with the mean for the Galactic thick disk from \cite{reddy06} 
indicated as a solid line.
The UMi giants follow the lower envelope of the halo field stars.
The data for Eu and other heavy neutron capture elements is discussed
in \S\ref{section_ncapture}.

\subsection{The Toy Model Fits of C09 Applied To the UMi Abundances \label{section_toy_model} }

The 10 UMi giants in our sample have [Fe/H] between $-1.35$
and $-3.10$~dex. To provide a context for the understanding
of our results we apply to the UMi sample the toy model fits for the behavior
of abundance ratios  [X/Fe] vs [Fe/H] developed for our sample in
the Draco dSph described in detail in C09.
This toy model was guided by the behavior of abundance ratios in Galactic
populations, the thin disk, thick disk, and Galactic halo field stars
since the same nucleosynthetic processes are involved, although they
may contribute different relative fractions to the chemical  inventory
in different environments.

Our toy model fits offer important clues for the importance
of various nucleosynthesis processes in the UMi and the Draco 
dSph galaxies as compared to
in the Galactic thick disk and halo stellar populations.
The parameters of the toy model depend on the nucleosynthetic
yields for the production channels for
each of the elements X and Fe,
the IMF, the rate of star formation,
accretion, loss of gas via galactic winds, interaction between the dSph
and the Milky Way via tides, ram pressure stripping, etc. as will
be discussed in \S\ref{section_chemev}.

The toy model sets [Fe/H](A) as the
mean for the  lowest metallicity stars in the UMi sample, 
and A(X) is the mean of [X/Fe] for
the same stars. [Fe/H](B) is the mean [Fe/H]
for the  highest metallicity stars, and B(X), a value of [X/Fe], is defined
similarly. The toy model 
represents such relationships as a plateau in [X/Fe] at the value [X/Fe](low)
over the range [Fe/H](A) to [Fe/H](low,X) and another plateau
at a value of [X/Fe](high), from [Fe/H](high,X) to [Fe/H](B).  A straight
line connects the two plateaus.
Thus our
model has four variables whose values are determined directly from the
dataset of [X/Fe] as a function of [Fe/H], with two additional
fit parameters.  We solve for the two free parameters in this toy model
[Fe/H](low,X) and [Fe/H](high,X)
by minimizing the variance around the fit.  The resulting parameters
are given in Table~\ref{table_fit}.

We apply this toy model to to 11 elements for which sufficient
accurate data is available for UMi members.  We use
the augmented sample of UMi giants, ignoring the outlier
COS171 which is discussed in detail in \S\ref{section_outlier}, 
leaving a sample of 15 UMi stars.
We use the two lowest metallicity stars in
the UMi sample to determine the plateau values A(X) and [Fe/H](A).
At the high metallicity end, we use 
three highest metallicity UMi stars,
each of which has [Fe/H] $-1.6{\pm}0.1$~dex obtained from a high quality
spectrum, to determine the plateau values B(X) and [Fe/H](B).
In solving for the two
fit parameters, weights
are halved for the 3 stars with lower accuracy spectra,
which are those from \cite{shetrone2} not reobserved with
the Subaru/HDS by \cite{subaru}.  
The  resulting parameters 
for each element
are listed in Table~\ref{table_fit} and the fits are shown
when available in Figs.~\ref{figure_nafeh} to \ref{figure_bafeh}.

The uncertainties in A(X) and in B(X) are approximately
those of $\sigma$[X/Fe]
for a single UMi star from our sample.  These values are given in
Tables~5 and 6 of C09.
Thus, for example, for [Mg/Fe] they are
$\pm$0.14~dex. A(Mg) is only 0.13~dex larger than B(Mg),
so 
the decline in [Mg/Fe] as [Fe/H] increases 
in the UMi sample is not statistically significant.
The decrease in [Na/Fe], [Si/Fe], and [Zn/Fe], 
and the increase in [Cr/Fe], [Mn/Fe]
and [Ba/Fe] as [Fe/H] increases are statistically significant.
Even when the change between the low and high metallicity abundance
ratio is clearly statistically significant,
the values for the knees of the distribution, [Fe/H](low)
and [Fe/H](high) are quite uncertain due to the small sample
of UMi giants coupled with the uncertainty of the individual
[X/Fe] determinations for each UMi giant.  

To overcome the large uncertainties in the location of the
knees of the fits, we have combined several elements,
assuming that at least some elements, if not all,
share the same values of [Fe/H](low,X)
and [Fe/H](high,X).   This
dramatically increases the number of
data points in the fit and lowers the uncertainties
for the final derived parameters.
 
\begin{figure}
\epsscale{1.0}
\plotone{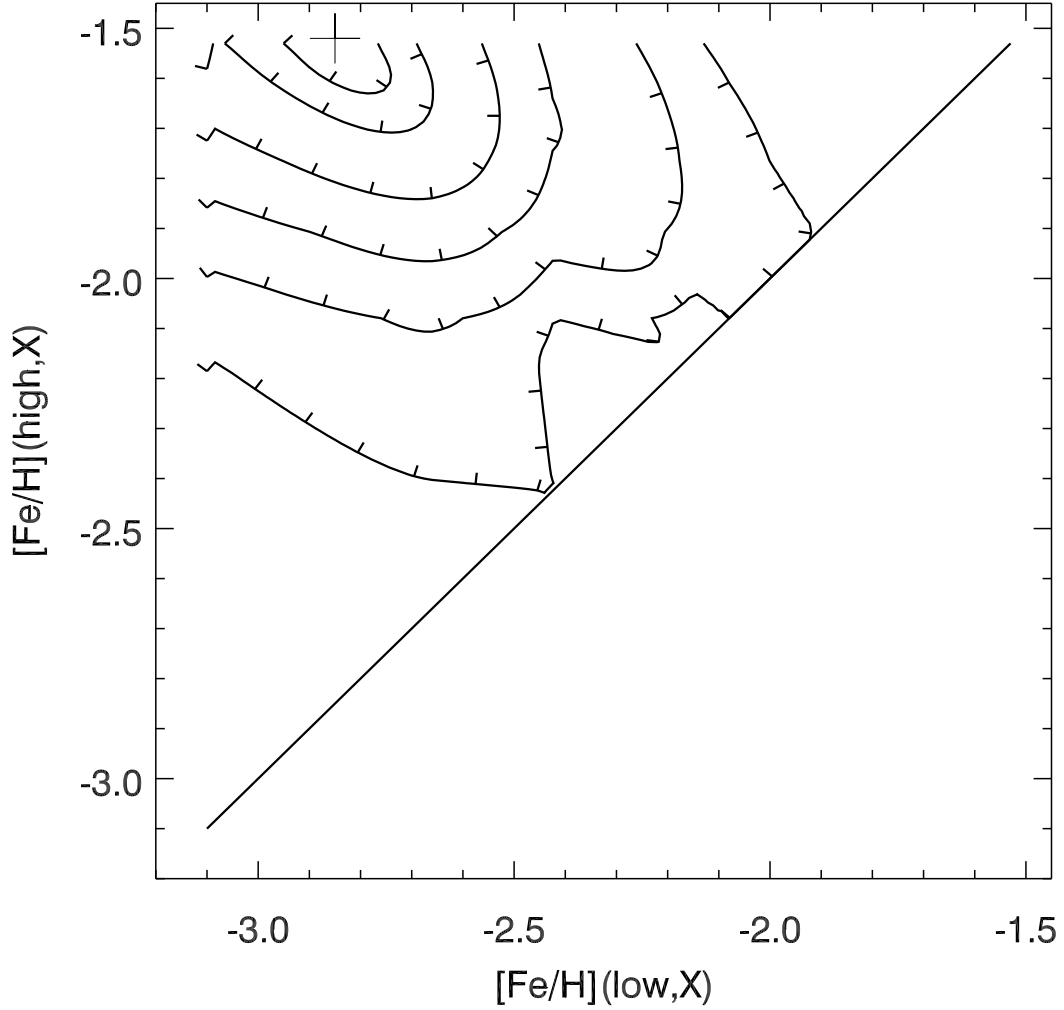}
\caption[]{
The contours of $\chi^2$ used to estimate the uncertainties
for [Fe/H](low,X) and [Fe/H](high,X) for the combined fit of the
ratios of Si, Cr, Mn, Ni, and Ba with respect to Fe as a 
function of [Fe/H] in the UMi sample.   The plus
sign shows where the $\chi^2_{\rm min}$ value is reached.
The contour lines are plotted around the minimum location
for $(N-3)(\chi^2-\chi^2_{\rm min})/\chi^2_{\rm min}=$1, 4,
16, 36, 64, and 100 respectively.  The short, perpendicular
ticks attached to each contour line show the``downhill''
direction of the $\chi^2$ ``valley''.
\label{figure_contour}}
\end{figure}

The [Fe/H](low/high,X) parameter space of interest is
limited to a triangular area in the
[Fe/H](low,X)-vs-[Fe/H](high,X) plane because
[Fe/H](low,X) $\leq$ [Fe/H](high,X).
In this area, 100$\times$100/2 sampling points
are uniformly distributed.  For each sampling point
(i.e. a pair of [Fe/H](low,X) and [Fe/H](high,X)),
we calculate the $\chi^2$ residual for each
element using the already determined values of A(X) and B(x)
for each element.  Then we add up the
residuals for all of the elements used in the combined
fit.  The summation is the $\chi^2$ residual
at that sampling point.  We apply this procedure
to all sampling points in the triangular area,
and obtain the residual $\chi^2$ valley for
the combined elements.  The lowest position of
the valley gives us the best fit parameters,
as is shown in Fig.~\ref{figure_contour}.
Three such combined fits were calculated; the results
are given as the last entries in
Table~\ref{table_fit}.

We estimate the uncertainties for the best combined
fit results as follows. We set $\chi^2_{\rm min}$
to the minimum value of all $\chi^2$ we calculated
within the triangular region of interest in the
[Fe/H](low,X)-vs-[Fe/H](high,X) plane.
Then the ``equal-altitude'' contour
line with $\chi^2-\chi^2_{\rm min}=\chi^2_{\rm min}/(N-3)$
roughly defines the 1-$\sigma$ range of the fitting
results, and that with
$\chi^2-\chi^2_{\rm min}=4\chi^2_{\rm min}/(N-3)$
roughly defines the 2-$\sigma$ range, where
$N$ is the number of data points used in
the fit procedure.   $N$ in a combined
fit is dramatically larger than in a single-element
fit. Each data point from \cite{shetrone2}
is counted as 0.5 in $N$, and their $\chi^2$ contribution
is also weighted by a factor of 0.5.  The derived
[Fe/H](high,X) for each of the three combined fits
is identical to within the uncertainties with
[Fe/H](B), suggesting we did not detect any plateau 
in [Fe/H] at the
high metallicity end of the UMi sample.

\subsection{The Heavy Neutron Capture Elements \label{section_ncapture} }

Just as the relative contribution of SNIa as compared to SNII to the
chemical inventory of the ISM provides a timescale, so too does
that of the $r$ versus the $s$-process for heavy neutron capture elements.
The $s$-process, reviewed in \cite{busso99}, 
occurs primarily in intermediate mass AGB stars.  The site of
the $r$-process is less clear, but 
is suspected to be in SNII during the formation of a neutron star
\citep{qian07}.

Here we emphasize the difference in behavior
for these elements between the UMi and Draco dSph galaxies
and the galactic halo field stars.  Unlike Draco, [Sr/Fe] (Fig.~\ref{figure_srfeh})
remains high (approximately at the solar ratio) over the full range
of [Fe/H] in UMi, while it falls for the lowest [Fe/H] stars in Draco.
The former behavior is that of the mean for the halo field,
while the latter is that of the low extreme of the Galactic  halo field population.
The  behavior of [Ba/Fe], though, is similar in the two dSph galaxies,
and lies at the low extreme of the range shown by halo field stars
at low metallicity.  Thus far, no star with [Ba/Fe] as low as the outlier
Draco~119 \citep{fulbright} has been found in UMi.  The low
outlier in UMi (COS171) is low for its rather high Fe-metallicity
for all the neutron capture elements; it is 
$\sim$0.8~dex low for [Ba, La, Ce, Nd, and 
Eu/Fe].

\begin{figure}
\epsscale{0.6}
\plotone{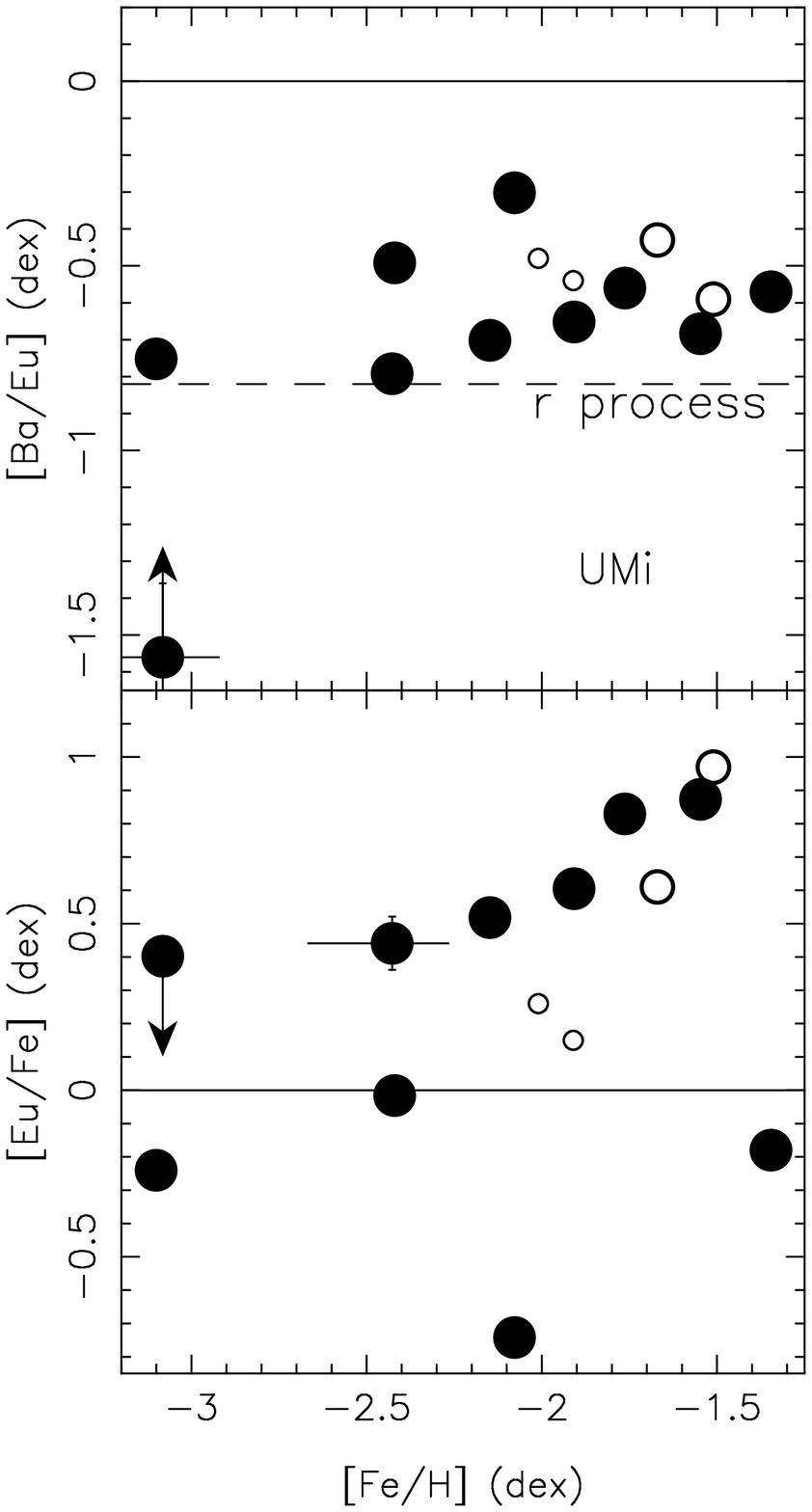}
\caption[]{[Ba/Eu] (upper panel) and [Eu/Fe] (lower panel) vs [Fe/H] 
is shown for  our UMi sample (large filled circles)
and the two from each of \cite{shetrone2} (small open circles) 
and \cite{subaru} (intermediate open
circles) with detected Eu.  
The Solar ratio is the solid horizontal
line, while the dashed horizontal line is the $r$-process
ratio from \cite{simmerer}; the $s$-process ratio, +1.4~dex, is above the top of
the figure.
Typical uncertainties are shown for one 
UMi star.
\label{figure_baeu}}
\end{figure}

But the most important difference between the behavior of UMi
and Draco for these elements is shown in the upper panel of Fig.~\ref{figure_baeu},
which displays the abundance ratio of the elements diagnostic for the
$s$-process (Ba) and the $r$-process (Eu).  
The solar $r$-process ratio shown in the top
panel is taken from
\cite{simmerer}; the solar ratio is a mixture of $r$ and $s$-process
material, while the pure $s$-process ratio for [Ba/Eu]
lies above the top
of the figure. 
Unlike Draco,
even at the highest metallicities reached in UMi,
 there is still
no sign of a contribution from the $s$-process, while in Draco
there is.   Another symptom of this is seen in the lower panel
of this figure, where the [Eu/Fe] ratios are very high for the
highest [Fe/H] giants in UMi, while
in C09 
they appear to drop towards solar for the higher metallicity Draco giants.
The exceptions in the lower panel are the outlier UMi~COS171 and UMi~28104,
which is strongly depleted in the neutron capture elements in the second (Ba) peak,
as is shown in Fig.~\ref{figure_bafeh}, but note that both stars
follow the rest of the UMi sample, displaying the
$r$-process ratio, in the upper panel of Fig.~\ref{figure_baeu}.

The UMi giants are slightly brighter than the Draco giants,
and this helps in the secure detection
of the many neutron-capture elements with only a few weak lines,
including La, Ce, and Nd,
among the stars at the metal-rich end of the UMi sample.
The ratios among these elements also support the conclusion
that the neutron capture elements in UMi originate
entirely in the $r$-process, as was suggested earlier
by \cite{subaru}.  As one expects in such a case, 
[Nd/Fe] shows an enhancement which is roughly 0.5 dex
smaller than that of [Eu/Fe], consistent with the
description of Nd as having roughly equal contributions
in the Sun from each of the $r$ and $s$-process, while 
for Eu, the $r$-process dominates.

\cite{simmerer} suggest that 
in the Galactic halo, signs of the $s$-process begin only at
[Fe/H] $> -2.6$~dex, and a mean [Eu/La] ratio halfway between
the pure $r$-process value and the Solar ratio is reached only
at [Fe/H] $\sim -1.4$~dex. The survey of cool metal-poor local dwarfs
of \cite{mashonkina} reaches the halfway
point in [Eu/Ba] from pure $r$-process to the Solar mixture only 
at [Fe/H] $\sim -0.5$~dex.  Thus the
result from Fig.~\ref{figure_baeu} 
is clear; the UMi distribution is close to that of the Galactic halo,
while in Draco, the $s$-process becomes important at 
a Fe-metallicity significantly lower than is characteristic
of the Galactic halo.

\begin{figure}
\epsscale{0.6}
\plotone{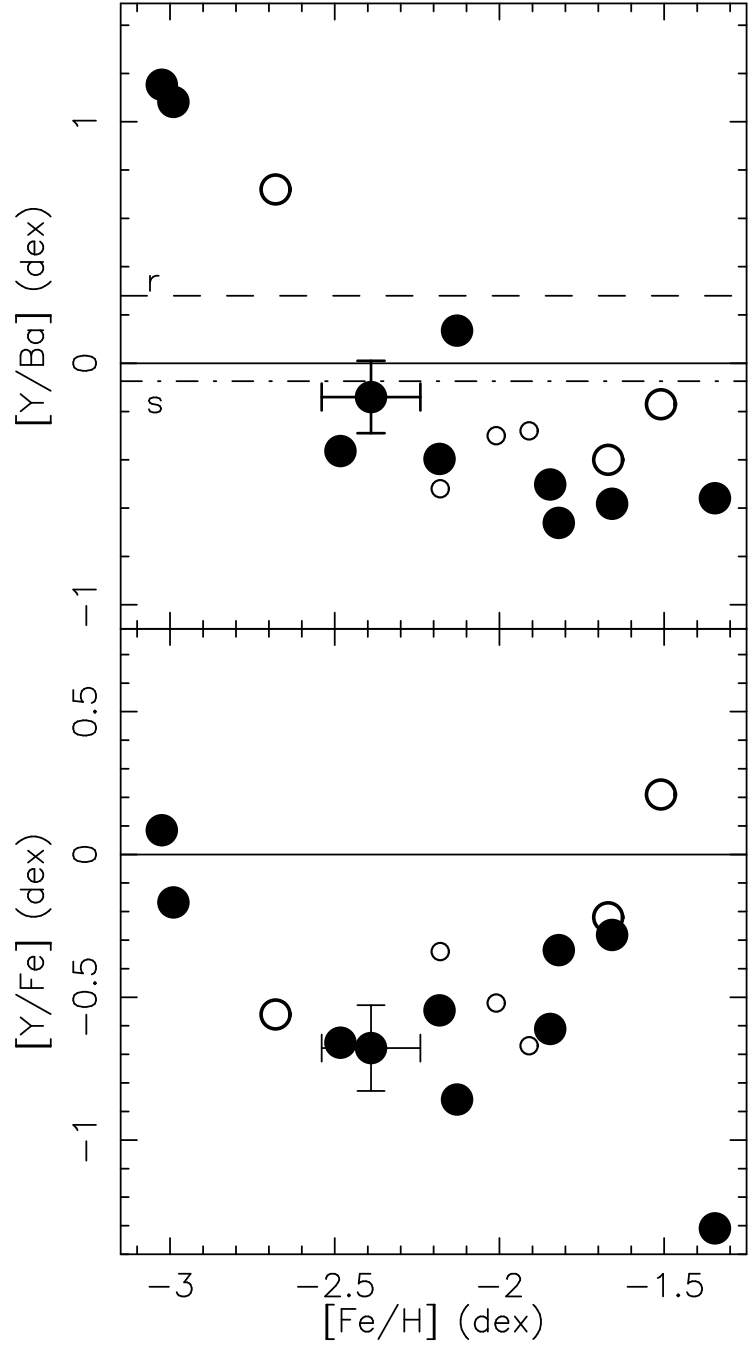}
\caption[]{
Upper panel: [Ba/Y] is shown as a function of [Fe/H] for our UMi sample.  The
symbols are those of Fig.~\ref{figure_baeu}.
The Solar ratio is the lower solid horizontal
line.  The pure $s$ and pure $r$-process ratios for the Sun
from \cite{simmerer} are indicated.  A typical error bar is shown
for one star.  Lower panel: [Y/Fe] vs [Fe/H] for the UMi sample.
\label{figure_bay}}
\end{figure}

The relative population of the first and second peaks
in neutron capture heavy element abundances is shown in  Fig.~\ref{figure_bay}
using Y vs Ba.  In the lowest metallicity UMi stars, one sees primarily
the very low fraction of Ba compared to a normal fraction of Y,
while in the somewhat higher metallicity UMi stars, [Ba/Fe]
approaches the solar value, and [Y/Ba] becomes slightly subsolar, and
well below the value typical of the halo 
field\footnote{The equivalent figure of C09, Fig.~19, shows
[Ba/Sr] as a function of [Fe/H].  The Draco data are correctly
plotted, but a mistake was made in the location of the
$r$ and $s$-process ratios in that figure.  They should both be
very close to the Solar ratio.}.
Clearly for the lowest metallicity UMi giants,
production of additional Sr and Y\footnote{The Zr abundances for
the UMi sample are quite uncertain.} by some additional mechanism 
such as the ``weak r-process''  or the
``lighter element primary process'' introduced by \cite{travaglio04}
is  required.

\subsection{Age -- Metallicity Relation for UMi \label{section_age}}

A very useful diagnostic of the star formation rate as a function
of time is the age -- metallicity relationship.  We construct this
for UMi using  [Fe/H] values  obtained  from detailed
abundance analyses for the augmented sample in UMi
including results from \cite{shetrone2}
and from \cite{subaru}.
[Ca/H] from \cite{winnick03} for the remaining members of UMi 
she observed at moderate dispersion
was transformed into [Ca/H](HIRES)
using using the  linear fit given in \S\ref{section_analysis}.
SDSS photometry from DR7 \citep{sdss_dr7} for these stars is combined with
the isochrones of \cite{dotter08}.  We adopt a relation between
[$\alpha$/Fe] and [Fe/H] based on our results described above.  
Given [Fe/H], [$\alpha$/Fe],
the colors, the distance of UMi and the adopted reddening, we can
determine the age of each UMi giant. 

\begin{figure}
\epsscale{0.9}
\plotone{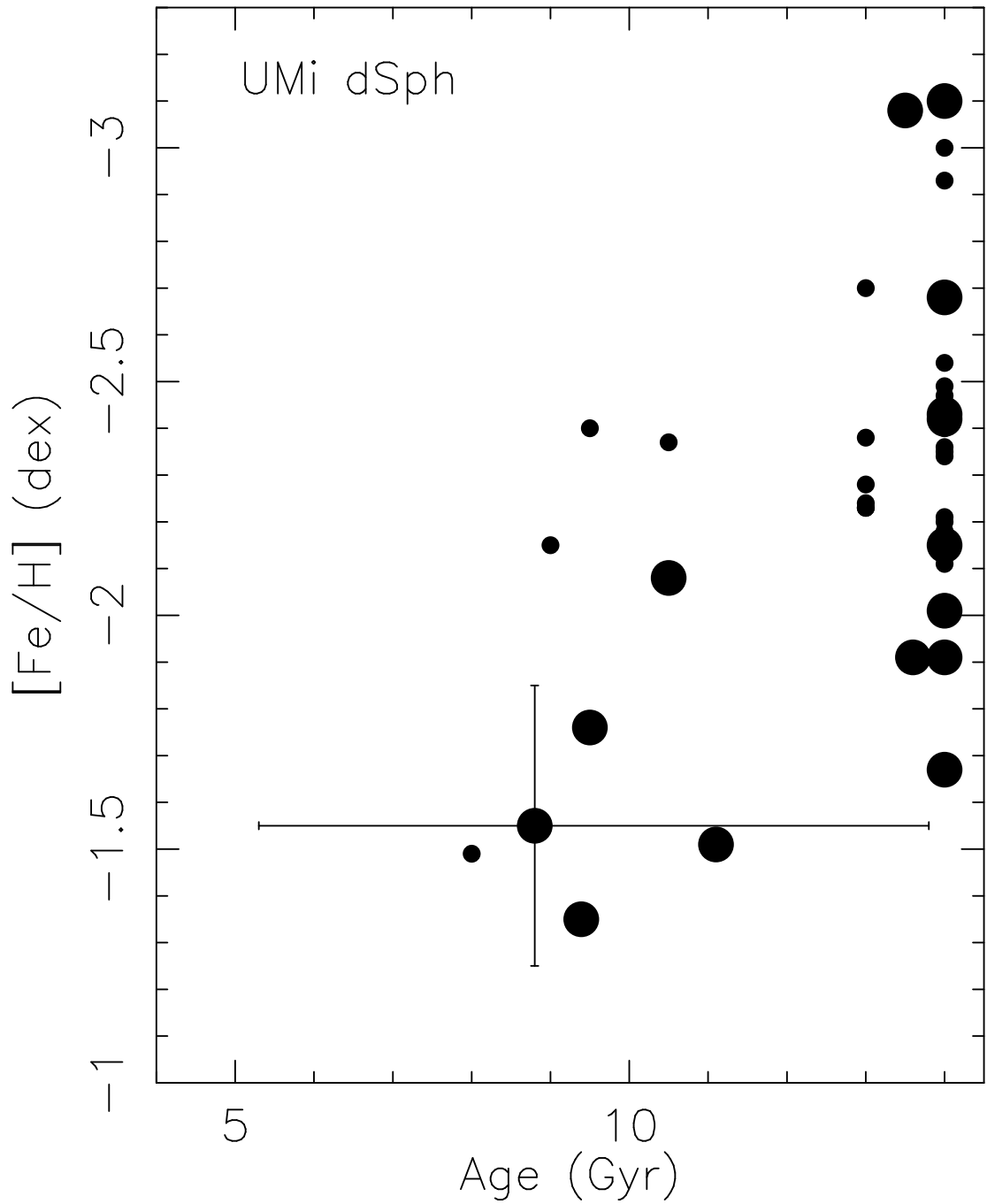}
\caption[]{The age of each giant with $M_i < -2.0$~mag
known to be a member of UMi  from Table~3.6 of \cite{winnick03}
is shown as a function of [Fe/H].
The Dartmouth isochrones \citep{dotter08} were used with
[Fe/H] from high resolution spectra or values derived from her near-IR 
Ca triplet measurements.  Typical error
bars are shown for a single star.
\label{figure_age_metal}}
\end{figure}

We do this for each star with $M_{i'} < -2.0$~mag.  
The isochrones along the RGB 
for lower luminosity stars converge too much in the ($g$' -- $i$') color
to attempt this.
The results are shown in 
Fig.~\ref{figure_age_metal}.  Stars which are slightly redder
than the reddest isochrone for the appropriate
metallicity are assigned ages of 14~Gyr.
The (large) uncertainty is these ages is discussed in C09.

The median age for the 40 UMi
giants is 14~Gyr, which is considerably higher than 
the median we found in C09 for Draco luminous giants. In the mean,
the UMi stars are of a uniform old age, in good agreement with the CMD
analysis of HST images by \cite{dolphin02} and more recently by \cite{orban08}.
There is a hint of an age-metallicity relation, with the highest metallicity 
stars being $\sim 3$~Gyr younger on average than the bulk of the 
UMi stellar population.

\subsection{Outliers in UMi and in Draco \label{section_outlier} }

As noted above, our UMi sample contains one outlier, COS171.
This star seems depleted in everything except Fe, or perhaps
received a substantial amount of pure Fe ejecta in addition
to a more normal mix.  In hindsight, Draco XI-2 may be a less
extreme case (see C09). The Galactic halo contains a 
very small number of very peculiar stars
\citep[see, e.g.][]{cohen08,lai_weird}, but none of
these known to the authors come close to matching the characteristics of COS171.
The low $\alpha$ stars discussed by \cite{ivans03}
show peculiarities only for the $\alpha$ and heavy
neutron capture elements and are 
are much less depleted in these elements with respect to
Fe than is COS171.  

While COS171 is indeed unique within the Draco, the UMi,
and Galactic halo samples
of stars with detailed abundance analyses, it is highly reminiscent
of the more extremely depleted stars in Fornax analyzed by
\cite{letarte07}, but COS171 has a somewhat lower [Fe/H].
This analogy holds through the Fe-peak, but not for the heavy
neutron capture elements.  Somehow this
star shares many of the characteristics of stars in a galaxy 
that has experienced
extended star formation over at least 5~Gyr
\citep{orban08} with a mean [Fe/H]
much higher than that of Draco or UMi.  It is interesting
to note that its estimated age (see \S\ref{section_age}) is  9.4~Gyr, 
considerably younger than
that of the vast majority of the UMi stellar population, but the
errors on this age are quite large.   

The only
outlier in the Draco sample discussed in C09,
Draco~119 \citep{fulbright}, has a very
different behavior; it is normal for most elements, and 
is a low outlier only  for
the neutron capture elements beyond the Fe-peak.  Given the low
star formation rate in these low-metallicity dSph galaxies
and the extremely low fraction of the neutron capture elements
even at Solar metallicity,
a wide range in the abundances of these very rare heavy elements, 
as is seen in  Galactic halo field stars at
very low metallicity, should be expected.

\section{Chemical Evolution of the UMI dSph Galaxy \label{section_chemev} }

In addition to our UMi analysis reported here and that of Draco in C09, as
of today, there are only three other dSph galaxies with published detailed
abundance analyses from high dispersion spectra
for 14 or more stars  to which we can compare
our Draco results.  These are the Sgr dSph (the main core, not the stream)
\citep{monaco05,sbordone07}
and the Carina dSph galaxy, for which
\cite{koch_carina}  combines his analysis of 10 giants
with 5 from the earlier study by \cite{shetrone03}.  The extensive
study of Fornax by \cite{letarte07} 
is not directly relevant as the lowest metallicity stars 
in their Fornax sample barely
overlap the highest metallicity giants in UMi or Draco.

In comparing our results for UMi, our earlier work in Draco (C09), and
published studies of more metal-rich dSph Galactic satellites,
there appears to be a pattern of moving from
a chemical inventory for dSph giants with [Fe/H] $\lesssim -2$~dex
which is very similar to that of stars in the
outer part of the Galactic halo (enhanced $\alpha$/Fe 
relative to  the Sun,
coupled with subsolar [X/Fe] for
the heavy neutron capture elements and $r$-process domination), 
switching to subsolar $\alpha$-elements
and super-solar  $s$-process dominated neutron capture elements
for the highest [Fe/H] dSph stars.  The combination of
low star formation rates over a varying and sometimes
extended duration that produced the stellar
populations in the local dSph galaxies with [Fe/H] $> -1.5$~dex
leads to a chemical inventory wildly discrepant from that of any
component of the Milky Way.

\begin{figure}
\epsscale{0.85}
\plotone{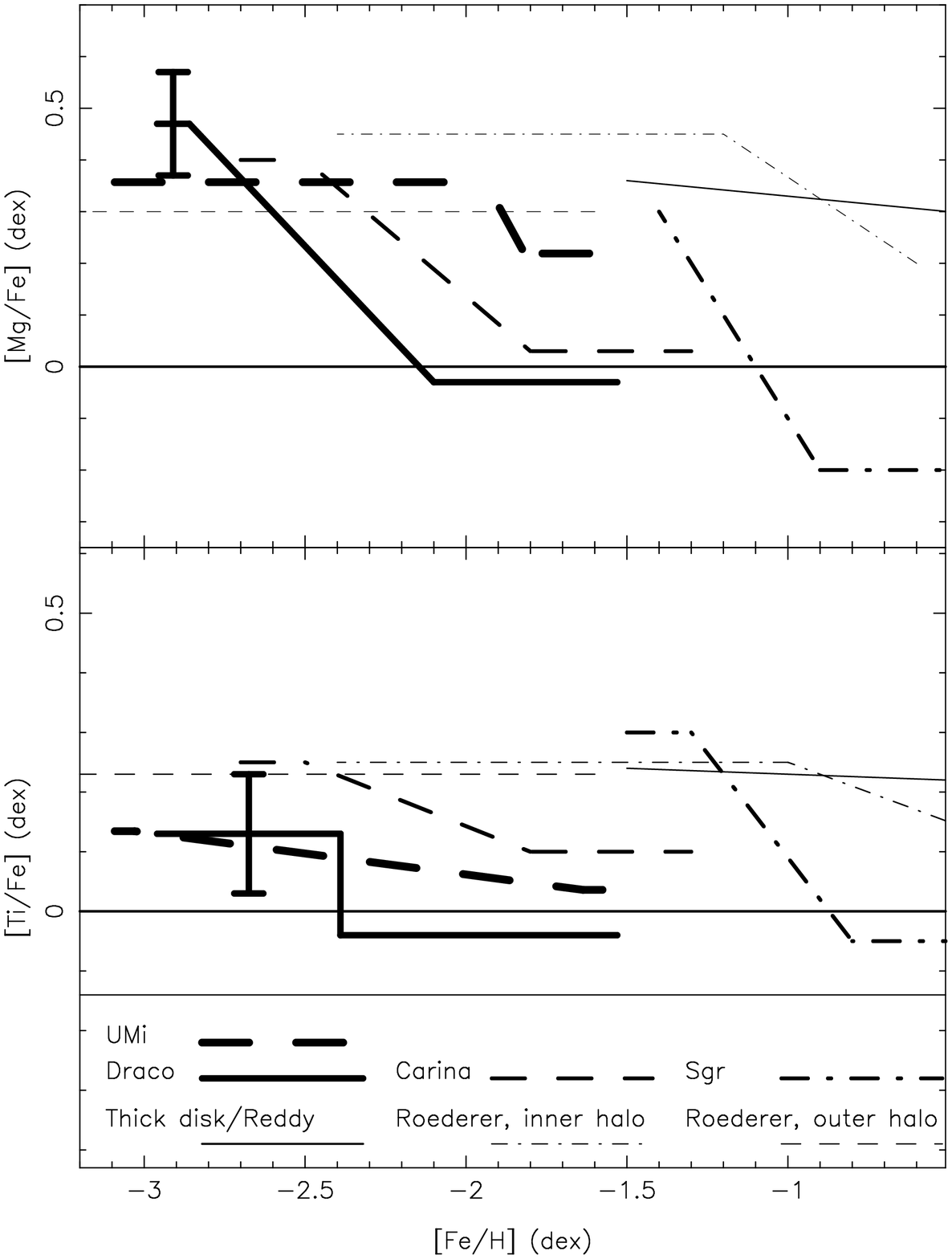}
\caption[]{
The toy model fit for [Mg/Fe] vs [Fe/H] (top panel)
and for [Ti/Fe] (bottom panel) for the sample
of 14 giants in each of the UMi and Draco dSph galaxies
is shown together with that for the Sgr 
\citep{monaco05,sbordone07}
and Carina \citep{koch_carina} dSph galaxies.  Fits for
Galactic components to data from \cite{roederer08} and from
\cite{reddy06} are shown as well.  Typical errors in abundance
ratios for the average of two stars are shown.
\label{figure_dsph_2panel}}
\end{figure}

To demonstrate this in detail,
we apply our toy model to the recent data for the Carina and the Sgr dSph
galaxies.  Fig.~\ref{figure_dsph_2panel} for [Mg/Fe] and 
for [Ti/Fe] show the fits for these two galaxies, for UMi, and for Draco. 
The fits for the Milky Way thin and thick disk are also displayed.
This figure clearly demonstrates the differences among the dSph satellites
for [Mg/Fe] as a function of [Fe/H].  The UMi sample is constant to within
the uncertainties
at [Mg/Fe] ~ $\sim ~ +0.35$~dex.  For the other
dSph galaxies
that relation begins high at low metallicity but [Mg/Fe] begins to decline at
considerably lower [Fe/H] for higher metallicity
star; the [Fe/H] at which this decline begins increases as the
mean metallicity of the dSph increases.  A similar situation
occurs for [Ti/Fe], but with a smaller total range in this ratio,
hence the differences in the trends for the various dSph galaxies
are less certain.
What definitely is changing between the various dSph galaxies 
is the Fe-metallicity range.
Among the dSph satellites with suitable abundance data,
the UMi and Draco systems have the the lowest mean [Fe/H] for their giants,
Carina is intermediate, and Sgr is closest 
to the Milky Way.

The knee values [Fe/H](X,low) and [Fe/H](X,high), which represent
the timescale (or,
more correctly, the [Fe/H])
at which the relative contributions of processed ejecta into the ISM
of  the system
from the various nucleosynthesis sites
change significantly, are also changing
for
these two (and other) elements among the various dSph Galactic satellites.
The Galactic thick and thin disk populations all approach [X/Fe] = 0 close to 
or at the
Solar Fe-metallicity, i.e. B(X) and [Fe/H](high,X) $\sim 0$. 
But in UMi and in Draco,
as is shown for UMi here in Table~\ref{table_fit} and for Draco in Table~8 of C09, 
the approach toward Solar ratios for some elements begins
at a considerably lower [Fe/H]. For example. 
[Fe/H](low,Si) is $-0.5$~dex for
the Milky Way thick disk, $-1.6$~dex for UMi,
and  $-2.0$~dex for Draco.
In addition, the 
relative contributions of the $r$ vs the $s$-process to the
production of heavy neutron capture elements varies
a lot among the dSph galaxies, with UMi showing
{\it{no}} detectable contribution from the $s$-process, believed to
originate from intermediate mass AGB stars,
consistent with its short epoch of star
formation. 

Recent models of chemical evolution for the disk, bulge, and halo of the Milky
Way based on the precepts first established by \cite{tinsley}
have been presented by several groups, including
\cite{timmes}, \cite{kobayashi}, \cite{prantzos}, and \cite{matteucci08}.
These models generally assume complete and uniform
mixing of the gas over the total volume considered at all times with
the exception of the more sophisticated model of 
\cite{marcolini06,marcolini08}.
Such models 
have been reasonably successful in reproducing the chemical
evolution of the major components of the Milky Way overall, although
failing in some (minor) details.

The evolution of the dSph galaxies differs in principle from
that of the Milky Way or its halo.  Their binding energies
are lower, so 
the importance of gas loss may be higher,
particularly in the case of material from SNII, for which the ejection velocity 
is  significantly larger than the escape velocity. 
Furthermore since
both UMi and Draco at present show no evidence for the presence of gas,
gas loss via a galactic wind or through interactions between the dSph satellite
and its host, the Milky Way, must have been important in the past. 
These galaxies also
show the consequences of lower star formation efficiency
which leads to slower star formation
overall without
the large initial burst that dominates nucleosynthesis in 
most of the Milky Way
components.  In a system where the star formation rate is slower and more
constant with time, SNIa ejecta can become important contributors
before [Fe/H] just from SNII builds up in the dSph interstellar medium
to  high values near $\sim -1$~dex.  It is this time delay
between the SNII and SNIa contributions that dominates discussion
of the chemical evolution of dSph galaxies.

\cite{lanfranchi04} suggest another
mechanism for affecting the $\alpha$/Fe ratios, namely the presence
of a strong outflow, which reduces the amount of gas available for
star formation. This in turn cuts off the production of $\alpha$-elements 
in massive
stars, while the SNIa rate, and the consequent production of Fe, 
continues unaffected.
This too could cause the drop in [$\alpha$/Fe] ratios 
common among the dSph galaxies.
Separating the contribution of a slow star formation versus a
strong outflow in the chemical history of a dSph is not easy
from abundance ratios alone.  It requires a knowledge of the 
metallicity distribution of the stars in the dSph, and ideally
of the age-metallicity distribution as well.  \cite{lanfranchi04}
claim that both effects are necessary to explain the
characteristics of UMi and of Draco.

\cite{matteucci08} reviews models for the chemical
evolution of the dSph galactic satellites of the Milky Way
that reproduce the behavior
of the $\alpha$-elements.  Presumably the agreement at
the lowest [Fe/H] values probed here, where the Galactic halo
stars overlap the Draco giants, is a consequence of 
a chemical inventory to which only SNII contributed,
but the trends in UMi are not as well reproduced.
\cite{lanfranchi04} present
detailed models for the evolution of
6 of the dSph Milky Way satellites, including UMi and Draco,
which try to reproduce not only the chemical evolution
but also the total stellar mass and  their individual star
formation histories as derived from CMD studies.
Their model for UMi has a very low
star formation efficiency and
the shortest duration of star formation (only 3~Gyr occurring immediately
after the galaxy condensed)
of these 6 dSph galaxies.  To within the uncertainties
of the measurements and the models, they succeed in reproducing the
almost flat  [Ca/Fe] relation with [Fe/H] of Fig.~\ref{figure_cafeh},
but their relation shows a fairly 
steep
decline in [Mg/Fe] vs [Fe/H] 
which is not seen in UMi (see Fig.~\ref{figure_mgfeh}).

There are a number of other problems when one compares
detailed chemical evolution models to our data.  \cite{lanfranchi08},
who address the production of heavy elements beyond the Fe peak
in dSph galaxies, substantially underpredict the ratio [Y/Fe]
for the most metal poor stars in UMi, and overpredict 
[Ba/Fe] for the same stars in both Draco and UMi.
The Ba/Eu ratio is predicted satisfactorily for these UMi stars,
but probably that is simply a result of the dominance
of the $r$-process in their production.
The cause of the relatively small difference 
in behavior of [Mg and Si/Fe] vs [Ca, and Ti/Fe] at the lowest
metallicities in UMi and in Draco is not clear, particularly
since Si is  an explosive $\alpha$-element
while Mg is a hydrostatic one.   How this  behavior relates to the mass
distribution of the SNII progenitors, given that one also
needs to reproduce the odd-even effect at [Sc/Fe], is not obvious.  
Qualitatively similar differences in the behavior
of the $\alpha$-elements vs [Fe/H] are also seen in the Galactic
bulge \citep{fulbright_bulge}, but again there are differences
in detail as the separation between hydrostatic and explosive
$\alpha$-elements is cleaner there, i.e. [Si/Fe] behaves like 
[Ca/Fe] and [Ti/Fe] in the Galactic bulge, which does not
appear to be the case for UMi.

\cite{carigi02} presented a
chemical evolution model for UMi which also requires a metal-rich
wind.  More sophisticated models for dSph galaxies are
presented by 
\cite{marcolini06} and by \cite{salvadori08}, who use a hierarchical merger tree
with a semi-analytical scheme galaxy formation.  These more complex but more
realistic models are rapidly improving but
are not yet fully capable of following chemical evolution
in detail.

\subsection{Implications for the Formation of the Galactic Halo 
\label{section_haloform} }

Whether the Galactic halo could have been formed by accretion of
satellite dwarf galaxies
has become a question of great current interest; see e.g. \cite{tolstoy03},
\cite{shetrone2}, among others.  Due largely to technical
advances and the construction of 8 to 10~m telescopes,
the 
data now available for the Galactic satellite galaxies
 is a tremendous improvement over that of a decade ago both in 
terms of number of stars analyzed and in accuracy of the results.
Recent efforts are summarized in the review by \cite{geisler08}.
The very recent review of \cite{tolstoy_araa}
focuses on their large ongoing project  at the VLT
to study dwarf galaxies  \citep[the DART project, ][]{tolstoy03}.

Our work in UMi and in Draco (see C09) and that published for the Carina
and for the Sgr dSph galaxies show that
abundance ratios among stars in dSph galaxies tend
to overlap those of Galactic halo giants at the lowest
Fe-metallicities probed.
This is only to be expected, as the nucleosynthesis ejecta from SNII
are to first order independent of metallicity. 
It is thus possible that the satellites were accreted early in their
development.  Their properties as we observe them today
would then not be relevant to this issue.
In 
each of these stellar populations, a minimum metallicity threshold
for formation of low mass stars along the lines of that
discussed by \cite{bromm} seems to exist.

\cite{helmi06} claimed  that
early accretion of satellites as a way of forming the Galactic halo
was still ruled out because 
of the metallicity distribution
function (MDF) they deduce for four dSph galaxies.
Given the MDF they used for the Galactic halo, they claimed that
dSph galaxies would be expected
to contain at least a few stars with [Fe/H] $< -3.0$~dex, while
they had not to date  detected any such stars in the four dSph galaxies
in which they have extensive samples from the DART 
project\footnote{\cite{starkenburg} very recently retracted these claims.}.

However we  found one such star in Draco (C09) and two more in UMi.
\cite{frebel_scl} have found a star at [Fe/H] $-3.8$~dex in the Sculptor
dSph galaxy.  \cite{aoki09} have found a star at $-3.10$~dex
in Sextans.  Several more such stars have been
found in the ultrafaint Milky Way satellites:
\cite{norris_boo} ([Fe/H] $-3.7$~dex in
Bootes I),  \cite{simon_leoiv}
([Fe/H] $-3.2$~dex in Leo IV), and \cite{frebel10}
(two giants, at $-3.10$ and $-3.23$~dex in UMa~II).   A number
of other stars found by \cite{kirby08} and \cite{kirby10}
in various dSph galaxies
are suspected to be below $-3.0$~dex, but most are too faint for
high-dispersion spectroscopy.  
 In addition \cite{hes_mdf} 
recently completed 
a determination of the halo MDF based
based on the Hamburg/ESO Survey  
which shows that completeness corrections are important in the
MDF derived from the HES. 

Collectively this very recent work
serves to help reestablish
the scenario for the formation of the Galactic halo via accretion
of satellite galaxies as viable.  The material
now  in the inner halo of the Galaxy had to have been
accreted early in the star formation history of the dSph galaxies,
giving time for orbital mixing to eliminate traces of discrete
stellar streams,
while satellite galaxies accreted somewhat later could contribute
to populating
the outer halo, which shares many of the abundance anomalies of
the dSph galaxies.  Dissolved globular clusters had to disperse
fairly quickly before the ubiquitous light element correlations
among O, Na, Mg and Al developed,
as these are not seen among halo field stars.

\section{Summary \label{section_summary}}

We present a detailed abundance analysis based on high resolution
spectra obtained with HIRES on the Keck~I Telescope
of 10 stars in the Ursa Minor dwarf spheroidal galaxy.
The sample was selected to span the full range in metallicity
inferred from \cite{winnick03}, who used moderate resolution spectroscopy
for radial velocity members of this dSph galaxy found by earlier surveys.
Her CaT indices of the strength of the near-infrared Ca triplet correlate 
well with [Ca/H] we derive from our detailed abundance analyses
with differences
from a linear fit of only $\sigma = 0.13$~dex. 

We use classical plane-parallel (1D) LTE models from the Kurucz grid
\citep{kurucz93} with a recent version of the stellar abundance
code MOOG \citep{moog}.
[Fe/H] for our sample stars ranges from $-1.35$ to $-3.10$~dex.
Combining our sample with previously published work
of \cite{shetrone2} for 6 UMi giants\footnote{We
ignore one carbon star from \cite{shetrone2}.},
and an analysis based on
higher S/N spectra of three UMi giants  by \cite{subaru}, two of which
were already studied in the earlier work,
gives a total
of 16 luminous UMi giants with detailed abundance analyses.

We find that for
the UMi sample [Mg/Fe] is constant to within the uncertainties  with
a value $\sim ~ +0.35$~dex for all the stars\footnote{The anomalous outlier
UMi COS171 is ignored here.  See \S\ref{section_outlier}.}, a trait
 shared by outer Galactic halo stars.
The abundance ratios  [Si/Fe], [Cr/Fe],
[Ni/Fe], [Zn/Fe], and perhaps [Na/Fe] and [Co/Fe] for the UMi giants
overlap those of
Galactic halo giants at the lowest [Fe/H] probed, but 
for the higher Fe-metallicity UMi stars are
significantly lower than those of Milky Way halo giants.
For the explosive $\alpha$-elements
Ca and Ti
the abundance ratios are also constant
to within the uncertainties but are somewhat low over the full metallicity range of
the UMi dSph stars compared to Galactic halo giants, being
closer, but still perhaps slightly low, at the lowest Fe-metallicities.
Nucleosynthetic
yields sensitive to the neutron excess, hence to the initial metallicity
of the SN progenitor \citep[see, e.g.][]{timmes}, may  be  
important in explaining the
origin of differences between UMi giants and Galactic field stars
for several of the abundance
ratios studied here. 

The heavy neutron capture elements in UMi giants
have $r$-process ratios at all
metallicities in UMi, consistent with the short duration
of its star forming epoch inferred from CMDs \citep{orban08}.
The relative contribution of these heavy elements seems to increase
as [Fe/H] increases for most of the UMi giants.

There are small, but real, differences between the trends
of abundance ratios between UMi and those of Draco from our
earlier study (see C09) which are discussed in detail in
\S\ref{section_field}.
There is one outlier in our UMi sample, which appears to have an excess of Fe,
or a depletion of essentially all elements with respect to Fe.
Similar behavior is seen among the most extreme of 
the Fornax dSph giants \citep{letarte07}.

In comparing our results for UMi, our earlier work in Draco (C09), and
published studies of more metal-rich dSph Galactic satellites,
there appears to be a pattern of moving from
a chemical inventory for dSph giants with [Fe/H] $\lesssim -2$~dex
which is very similar to that of stars in the
outer part of the Galactic halo
(enhanced $\alpha$/Fe 
relative to  the Sun,
coupled with subsolar [X/Fe] for
the heavy neutron capture elements and $r$-process domination),
switching to subsolar $\alpha$-elements
and super-solar  $s$-process dominated neutron capture elements
for the highest [Fe/H] dSph stars.  The combination of
low star formation rates over a varying and sometimes
extended duration that produced the stellar
populations in the local dSph galaxies with [Fe/H] $> -1.5$~dex
leads to a chemical inventory wildly discrepant from those seen in any
component of the Milky Way.

The dominant uncertainty in these results is the possibility of differential
non-LTE or 3D effects 
between the very cool luminous giants in our sample from the
UMi and Draco dSph galaxies 
and the comparison halo field and globular cluster stars,
which are somewhat hotter.  With a 30-m telescope it will be possible
to reach lower luminosity and somewhat hotter giants in the dSph
satellites of the Milky Way where these
issues will be less important. 

In C09 we developed a toy model fit which we use to illuminate these trends, and to
compare them with those of Galactic globular clusters and of
giants from the Carina and Sgr dSph galaxies.   Since there is  
good agreement in most
cases for the abundance ratios at the lowest metallicity within
a given sample and also the highest metallicities sampled, the fundamental
contributors to their chemical inventory 
(SNII at the lowest
 metallicity and SNIa plus other sources at the highest [Fe/H])
behave in very similar ways
 in all these environments.  We thus infer that the IMF for massive stars must be
 similar as well.
The key differences lie in the [Fe/H] 
corresponding to the knee values, 
i.e. in the timescale (or,
more correctly, the [Fe/H])
at which the relative contributions of processed ejecta into the ISM
of  the system
from the various nucleosynthesis sites
change significantly.
The UMi and Draco systems, which have among the lowest luminosities 
for the classical dSph satellites of the Milky Way, have the lowest mean 
[Fe/H] for its giants,
Sgr is intermediate, and  the Carina dSph is closest 
to the Milky Way  halo and the 
thick disk.    Our new data will enable
much more sophisticated modelling of the chemical evolution
of Draco with more detail than our simple toy model can provide.

We note the
presence of two luminous giants in our UMi sample 
with [Fe/H] $< -3.0$~dex.
This combined with other recent evidence for a small number
of extremely low metallicity stars in other dSph galaxies
reaffirms that the inner Galactic halo could have largely been formed
by early accretion and dissolution of Galactic satellite galaxies and by
globular clusters which dissolved prior to the imprinting of an AGB signature,
while the outer halo could have formed largely 
from those dSph galaxies  accreted later. 

The age--metallicity relationship established by combining
photometry, spectroscopic metallicities, and isochrones
suggests that the stellar population in UMi consists of
old metal-poor stars.  
Unlike our previous result in C09 for Draco,
there is no evidence for
the presence of an intermediate age component in UMi.
There is a hint of an age-metallicity relationship with
the most metal-rich UMi stars being $\sim 3$~Gyr younger
than the metal-poor old population.

\acknowledgements

The entire Keck/HIRES and LRIS user communities owes a huge debt to
Jerry Nelson, Gerry Smith, Steve Vogt, and many other
people who have worked to make the
Keck Telescope and HIRES a reality and to operate and
maintain the Keck Observatory. We are grateful to the
W. M.  Keck Foundation for the vision to fund
the construction of the W. M. Keck Observatory.  The authors wish 
to extend
special thanks to those of Hawaiian ancestry on whose sacred mountain
we are privileged to be guests.  Without their generous hospitality,
none of the observations presented herein would
have been possible.

The authors are grateful to NSF grant AST-0507219
and grant AST-0908139  for partial support.
This publication makes use of data from the Two Micron All-Sky Survey,
which is a joint project of the University of Massachusetts and the 
Infrared Processing and Analysis Center, funded by the 
National Aeronautics and Space Administration and the
National Science Foundation.

{}

\clearpage

\begin{deluxetable}{rccccc c}
\tablewidth{0pc}
\tabletypesize{\scriptsize}
\tablecaption{The Sample of Stars in the Ursa Minor dSph
\label{table_sample}}
\tablehead{
\colhead{ ID\tablenotemark{a} } &
\colhead{Coords.\tablenotemark{a} } &
\colhead{V\tablenotemark{a} } &
\colhead{I\tablenotemark{a} } &
\colhead{Date Obs. } &
\colhead{Exp. Time } & \colhead{S/N\tablenotemark{b}} \\
\colhead{  } &
\colhead{(J2000) } &
\colhead{(mag) } &
\colhead{(mag) } &
\colhead{ } &
\colhead{(sec)} & \colhead{ } }
\startdata
COS171 &  15 08 04.97~~~$+$67 14 00.8\phn  &  17.25 & 15.94  & 
    06/2008 &     7200 & 80 \\ 
COS233 & 15 08 10.01~~~$+$67 17 24.5\phn  &  16.93 & 15.61  &  
    06/2008 &  7200 & 95 \\
JI2 & 15 10 27.11~~~$+$67 24 36.2\phn & 16.98 & 15.65 & 06/2008 &
   7200 & 80\\
JI19    & 15 10 11.68~~~$+$67 08 28.8\phn  &  17.26 & 16.07  &  
     06/2008 &  7200 & 80\\
N37 & 15 11 05.76~~~$+$67 13 15.3\phn  & 17.07 & 15.71 & 06/2008 &
   7200 & 95\\
27940 & 15 11 24.30~~~$+$67 33 50.2\phn & 16.81 & 15.41 & 09/2009 & 
   6000 & 85 \\
28104 &  15 13 12.33~~~$+$67 33 20.6\phn &  16.86 & 15.50 & 
   02/2010 & 8100 & 105 \\   
33533   & 15 13 26.09~~~$+$67 16 22.7\phn &  16.90 & 15.62  & 
 06/2008 &  7200 & 100 \\
36886 &  15 13 30.30~~~$+$67 06 37.3\phn & 17.01 & 15.76 & 
   02/2010 & 7400 &  80 \\
41065   &  15 11 01.48~~~$+$66 52 54.5\phn  &  16.71 & 15.35  &  
  09/2009 &  4900 & 85 \\
\enddata
\tablenotetext{a}{The star names, coordinates, and photometry are from 
\cite{winnick03}, with the original source of the
latter two being \cite{palma03}. }
\tablenotetext{b}{S/N per spectral resolution element 
($\lambda/\Delta \lambda = 35,000$) in the continuum at the center of the 
echelle order at 5800~\AA.}
\end{deluxetable}

\begin{deluxetable}{rcc cc cr}
\tablewidth{0pc}
\tablecaption{Stellar Parameters for the Ursa Minor Giants \label{table_param}}
\tablehead{
\colhead{ ID } &
\colhead{$T_{\rm eff}$ } &
\colhead{$\log (g)$ } &
\colhead{$T_{\rm eff}$ } &
\colhead{$\log (g)$ } &
\colhead{$v_t$ } & \colhead{$v_r$} \\
\colhead{} &
\colhead{(K) (phot) } &
\colhead{(dex) (phot) } &
\colhead{(K) (spec) } &
\colhead{(dex) (spec) } &
\colhead{(km~s$^{-1}$)} & \colhead{(km~s$^{-1}$)} }
\startdata
COS171 &  4290 & 0.80 & 4380 & 0.8 & 1.9 & $-256.2$ \\ 
COS233 &  4425 & 0.80 & 4425 & 0.8 & 2.2 & $-266.3$ \\
JI2 & 4415 & 0.80 & 4415 & 0.85 & 2.1 & $-254.7$ \\
JI19  &  4530 & 1.0 & 4450 & 0.8 & 3.0 & $-246.2$ \\
N37 & 4290 & 0.80 & 4390 & 0.8 & 1.8 &    $-240.4$ \\
27940 & 4290 & 0.70 &  4290 & 0.7 & 2.2 & $-248.4$ \\
28104 & 4365 & 0.65 &  4365 & 0.65 & 2.3 & $-244.2$ \\
33533 &  4525 & 0.80 &  4450 & 0.85 & 3.0 & $-248.9$ \\
36886 & 4400 & 0.75 & 4400  & 0.75 & 2.6 & $-231.0$ \\
41065 &  4400 & 0.60 &  4350 & 0.6 & 2.6 & $-259.7$ \\
\enddata
\end{deluxetable}

\clearpage

\begin{deluxetable}{lccr rrrrr rrrrr}
\tabletypesize{\scriptsize}
\tablewidth{0pc}
\tablecaption{Equivalent Widths for Ten Stars in the Ursa Minor dSph
\label{table_eqw}}
\tablehead{
\colhead{$\lambda$} &
\colhead{Ion} &
\colhead{$\chi$ } &
\colhead{$\log (gf)$ } &
\colhead{33533 } &
\colhead{JI19 } &
\colhead{36886} &
\colhead{41065 } &
\colhead{COS233} &
\colhead{28104} &
\colhead{27940} &
\colhead{JI2} &
\colhead{N37} &
\colhead{COS171} \\
\colhead{\AA} &
\colhead{ } &
\colhead{(eV) } &
\colhead{(dex) } &
\colhead{(m\AA) } &
\colhead{(m\AA) } &
\colhead{(m\AA) } &
\colhead{(m\AA) } &
\colhead{(m\AA) } &
\colhead{(m\AA) } &
\colhead{(m\AA) } &
\colhead{(m\AA) } &
\colhead{(m\AA) } &
\colhead{(m\AA) }}
\startdata
 6300.30 & [O~I] &   0.00 &  $-$9.780 & \nodata & \nodata &   17.9 &   10.0 & \nodata & \nodata & \nodata & \nodata &   31.0 &  \nodata \\
 6363.78 & [O~I] &   0.02 & $-$10.300 & \nodata & \nodata & \nodata &    8.0 &    8.1 &    8.8 &   10.0 &    8.5 &    8.1 &    12.6 \\
 7771.94 & O &   9.15 &   0.369 & \nodata & \nodata & \nodata & \nodata & \nodata & \nodata & \nodata & \nodata & \nodata &     8.5 \\
 5682.63 & Na~I  &   2.10 &  $-$0.700 & \nodata & \nodata & \nodata & \nodata & \nodata & \nodata &    9.4 &    8.0 &   13.5 &  \nodata \\
 5688.19 & Na~I  &   2.10 &  $-$0.420 & \nodata & \nodata & \nodata &    9.8 &   12.0 &   10.0 &   15.3 &   15.5 &   22.5 &    12.0 \\
 5889.95 & Na~I  &   0.00 &   0.110 &  150.3 &  161.3 &  224.4 &  225.6 &  236.1 &  269.7 &  300.9 &  240.5 &  255.0 &   218.1 \\
 5895.92 & Na~I  &   0.00 &  $-$0.190 &  121.7 &  134.4 &  189.5 &  207.7 &  210.7 &  220.4 &  259.6 &  214.8 &  225.4 &   205.1 \\
 4703.00 & Mg~I  &   4.34 &  $-$0.440 &   43.0 &   53.6 &   85.8 &  113.0 &  115.0 &  124.1 &  128.5 &  118.2 &  136.3 &   107.3 \\
\enddata 
\tablenotetext{a}{Dubious line, probably should not have been used.}
\tablenotetext{b}{This line gives an abundance $\sim$1.0~dex higher than the
other two Dy~II lines, and is ignored in computing the Dy abundance.}
\end{deluxetable}

\clearpage

\begin{deluxetable}{c rrr}
\tablewidth{0pc}
\tablecaption{Fit Fe~I Slopes With EP, Equivalent Width, and Wavelength 
\label{table_slopes}}
\tablehead{\colhead{Star ID} &
\colhead{$\Delta$[X/Fe]/$\Delta$(EP)\tablenotemark{a}} &
\colhead{$\Delta$[X/Fe]/${\Delta}[W_{\lambda}/\lambda]$} &
\colhead{$\Delta$[X/Fe]/${\Delta}\lambda$} \\ 
\colhead{} & \colhead{(dex/eV)} & \colhead{(dex)} & 
\colhead{($10^{-4}$dex/$\AA$)}
}
\startdata
COS171 &  0.00   & $-0.03$ & $-0.07$ \\
COS233 & $-0.05$ & 0.01 & $-0.09$ \\
JI2    & $-0.04$ & 0.00 & $-0.18$ \\
JI19   & $-0.07$ & 0.03 & $-0.05$ \\
N37    & $-0.04$ & 0.00 & $-0.59$ \\
27940  & $-0.02$ & $-0.02$ & $-0.43$ \\
28104  & $-0.05$ & $-0.02$ & $-0.25$ \\
36886  & $-0.05$ & $-0.03$ & $-0.29$ \\
33533  & $-0.11$\tablenotemark{b} & 0.01 & 0.04 \\
41065  & $-0.06$ & $-0.02$ & $-0.23$ \\
\enddata
\tablenotetext{a}{Typical range of EP is 4 eV.
This slope decreases by $\sim$0.1~dex/eV for an  increase
in \teff\ of 250~K.}
\tablenotetext{b}{Correlation coefficient is very low.}
\end{deluxetable}

\clearpage

\clearpage

\begin{deluxetable}{l | rrrr | rrrr | rrrr | rrrr  | rrrr }
\tablenum{5a}
\tabletypesize{\tiny}
\rotate
\tablewidth{0pt}
\setlength{\tabcolsep}{.1cm}
\tablecaption{Abundances for the First Five UMi Stars\label{table_abunda}}
\tablehead{\colhead{Species} & 
\colhead{} & \multispan{3}{UMi~33533~~~[Fe/H] $-$3.10} &
\colhead{} & \multispan{3}{UMi~JI19~~~[Fe/H] $-$3.08} &
\colhead{} & \multispan{3}{UMi~36886~~~[Fe/H] $-$2.43} &
\colhead{} & \multispan{3}{UMI~41065~~~[Fe/H] $-$2.42} &
\colhead{} & \multispan{3}{UMi~COS233~~~[Fe/H] $-$2.15}   \\
\colhead{} & 
\colhead{[X/Fe]} & \colhead{log$\epsilon(X)$} & \colhead{No.} & 
       \colhead{$\sigma$\tablenotemark{a}} &  
\colhead{[X/Fe]} & \colhead{log$\epsilon(X)$} & \colhead{No.} & 
       \colhead{$\sigma$\tablenotemark{a}} &         
\colhead{[X/Fe]} & \colhead{log$\epsilon(X)$} & \colhead{No.} & 
    \colhead{$\sigma$\tablenotemark{a}} &
\colhead{[X/Fe]} & \colhead{log$\epsilon(X)$} & \colhead{No.} & 
    \colhead{$\sigma$\tablenotemark{a}} &
\colhead{[X/Fe]} & \colhead{log$\epsilon(X)$} & \colhead{No.} & 
    \colhead{$\sigma$\tablenotemark{a}} \\
\colhead{} & \colhead{(dex)} & \colhead{(dex)} & \colhead{Lines} & \colhead{(dex)} &
\colhead{(dex)} & \colhead{(dex)} & \colhead{Lines} & \colhead{(dex)} &
\colhead{(dex)} & \colhead{(dex)} & \colhead{Lines} & \colhead{(dex)} &
\colhead{(dex)} & \colhead{(dex)} & \colhead{Lines} & \colhead{(dex)} &
\colhead{(dex)} & \colhead{(dex)} & \colhead{Lines} & \colhead{(dex)} 
}
\startdata 
C(CH)\tablenotemark{b}  &  $-$0.03 &   5.46   &    1 &  \nodata   & 
    $-$0.15 &   5.36   &    1 &  \nodata   &  $-$0.88 &   5.28   &    1 &  \nodata   &  $-$0.66 &   5.51   &    1 &  \nodata   &  $-$0.76 &   5.68   &    1 &  \nodata   \\
O~I    & \nodata & \nodata   &  \nodata &  \nodata   & \nodata & \nodata   &  \nodata &  \nodata   &   0.58 &   6.99   &    1 &  \nodata   &   0.39 &   6.80   &    1 &  \nodata   &   0.56 &   7.24   &    1 &  \nodata   \\
Na~I   &  $-$0.42 &   2.80   &    2 &    0.06   &  $-$0.23 &   3.01   &    2 &    0.06   &  $-$0.17 &   3.73   &    2 &    0.06   &  $-$0.29 &   3.61   &    3 &    0.15   &  $-$0.37 &   3.80   &    3 &    0.17   \\
Mg~I   &   0.28 &   4.72   &    4 &    0.13   &   0.43 &   4.89   &    4 &    0.12   &   0.42 &   5.53   &    5 &    0.05   &   0.42 &   5.54   &    5 &    0.21   &   0.34 &   5.73   &    5 &    0.19   \\
Al~I\tablenotemark{c}    &   0.53 &   3.90   &    1 &  \nodata   & 
   0.47 &   3.85   &    1 &  \nodata   &  $-$0.53 &   3.51   &    1 &  \nodata   & \nodata & \nodata   &  \nodata &  \nodata   &  $-$0.18 &   4.15   &    1 &  \nodata   \\
Si~I   &   0.36 &   4.81   &    1 &  \nodata   &   0.98 &   5.45   &    2 &    0.10   &   0.34 &   5.47   &    5 &    0.17   &   0.44 &   5.57   &    4 &    0.24   &   0.29 &   5.69   &    4 &    0.16   \\
K~I    &   0.04 &   2.06   &    1 &  \nodata   &   0.45 &   2.49   &    1 &  \nodata   &   0.40 &   3.09   &    1 &  \nodata   &   0.22 &   2.92   &    1 &  \nodata   &   0.35 &   3.32   &    1 &  \nodata   \\
Ca~I   &   0.05 &   3.31   &    8 &    0.19   &   0.33 &   3.61   &   12 &    0.19   &   0.14 &   4.07   &   18 &    0.15   &   0.11 &   4.05   &   16 &    0.16   &   0.05 &   4.26   &   19 &    0.15   \\
\enddata
\tablenotetext{a}{$\sigma$ is the dispersion of the abundance ratio 
for each absorption line 
   about the mean for the species.}
\tablenotetext{b}{The 4320~\AA\ region of the G band of CH was used.}
\tablenotetext{c}{A non-LTE correction of +0.6~dex was used for stars where
only the 3961~\AA\ resonance line was measured.}
\tablenotetext{d}{An offset of +0.2~dex was applied if only
the 4030~\AA\ triplet lines could be detected.}
\tablenotetext{e}{[Fe/H](Fe~I) is given instead of [X/Fe].}
\end{deluxetable}

\clearpage

\begin{deluxetable}{l | rrrr | rrrr | rrrr | rrrr | rrrr }
\tablenum{5b}
\tabletypesize{\tiny}
\rotate
\tablewidth{0pt}
\setlength{\tabcolsep}{.1cm}
\tablecaption{Abundances for the Last Five UMi Stars\label{table_abundb}}
\tablehead{\colhead{Species} & 
\colhead{} & \multispan{3}{UMi~28104~~~[Fe/H] $-$2.08} &
\colhead{} & \multispan{3}{UMi~27940~~~[Fe/H] $-$1.91} &
\colhead{} & \multispan{3}{UMi~JI2~~~[Fe/H] $-$1.76} &
\colhead{} & \multispan{3}{UMi~N37~~~[Fe/H] $-$1.55} &
\colhead{} & \multispan{3}{UMi~COS171~~~[Fe/H] $-$1.35}   \\
\colhead{} & 
\colhead{[X/Fe]} & \colhead{log$\epsilon(X)$} & \colhead{No.} & 
       \colhead{$\sigma$\tablenotemark{a}} &  
\colhead{[X/Fe]} & \colhead{log$\epsilon(X)$} & \colhead{No.} & 
    \colhead{$\sigma$\tablenotemark{a}} &
\colhead{[X/Fe]} & \colhead{log$\epsilon(X)$} & \colhead{No.} & 
    \colhead{$\sigma$\tablenotemark{a}} &    
\colhead{[X/Fe]} & \colhead{log$\epsilon(X)$} & \colhead{No.} & 
    \colhead{$\sigma$\tablenotemark{a}} &
\colhead{[X/Fe]} & \colhead{log$\epsilon(X)$} & \colhead{No.} & 
    \colhead{$\sigma$\tablenotemark{a}} \\
\colhead{} & \colhead{(dex)} & \colhead{(dex)} & \colhead{Lines} & \colhead{(dex)} &
\colhead{(dex)} & \colhead{(dex)} & \colhead{Lines} & \colhead{(dex)} &
\colhead{(dex)} & \colhead{(dex)} & \colhead{Lines} & \colhead{(dex)} &
\colhead{(dex)} & \colhead{(dex)} & \colhead{Lines} & \colhead{(dex)} &
\colhead{(dex)} & \colhead{(dex)} & \colhead{Lines} & \colhead{(dex)} 
}
\startdata 
C(CH)\tablenotemark{b}  &  $-$0.54 &   5.97   &    1 &  \nodata   &
      $-$0.82 &   5.86   &    1 &  \nodata   &  $-$0.77 &   6.06   &    1 &  \nodata   &  $-$0.68 &   6.36   &    1 &  \nodata   &  $-$1.06 &   6.18   &    1 &  \nodata   \\
O~I    &   0.44 &   7.20   &    1 &  \nodata   &   0.34 &   7.26   &    1 &  \nodata   &   0.20 &   7.26   &    1 &  \nodata   &   0.15 &   7.43   &    2 &    0.09   &   0.02 &   7.51   &    2 &    0.24   \\
Na~I   &  $-$0.40 &   3.84   &    3 &    0.12   &  $-$0.44 &   3.97   &    4 &    0.06   &  $-$0.64 &   3.92   &    4 &    0.03   &  $-$0.72 &   4.05   &    4 &    0.10   &  $-$1.27 &   3.70   &    3 &    0.08   \\
Mg~I   &   0.40 &   5.87   &    5 &    0.19   &   0.30 &   5.93   &    6 &    0.15   &   0.23 &   6.00   &    4 &    0.12   &   0.13 &   6.12   &    4 &    0.18   &  $-$0.35 &   5.85   &    3 &    0.22   \\
Al~I\tablenotemark{c}   &   0.13 &   4.52   &    1 &  \nodata   & 
   \nodata & \nodata   &  \nodata &  \nodata   & \nodata & \nodata   &  \nodata &  \nodata   &   0.01 &   4.93   &    2 &    0.21   & \nodata & \nodata   &  \nodata &  \nodata   \\
Si~I   &   0.21 &   5.68   &    6 &    0.16   &   0.12 &   5.77   &   10 &    0.12   &   0.14 &   5.93   &    9 &    0.18   &   0.04 &   6.04   &    9 &    0.10   &  $-$0.19 &   6.01   &    9 &    0.15   \\
K~I    &   0.24 &   3.29   &    1 &  \nodata   &   0.26 &   3.47   &    1 &  \nodata   &   0.16 &   3.52   &    1 &  \nodata   &   0.20 &   3.78   &    1 &  \nodata   &  $-$0.12 &   3.65   &    1 &  \nodata   \\
Ca~I   &   0.02 &   4.30   &   20 &    0.14   &   0.01 &   4.46   &   17 &    0.21   &  $-$0.08 &   4.52   &   18 &    0.18   &  $-$0.10 &   4.71   &   17 &    0.16   &  $-$0.36 &   4.66   &   18 &    0.18   \\
\enddata
\tablenotetext{a}{$\sigma$ is the dispersion of the abundance ratio 
for each absorption line 
   about the mean for the species.}
\tablenotetext{b}{The 4320~\AA\ region of the G band of CH was used.}
\tablenotetext{c}{A non-LTE correction of +0.6~dex was used for  stars where
only the 3961~\AA\ resonance line was measured.}
\tablenotetext{d}{[Fe/H](Fe~I) is given instead of [X/Fe].}
\end{deluxetable}

\clearpage

\begin{deluxetable}{l rr rr rr}
\tablenum{6}
\tablewidth{0pt}
\tablecaption{Parameters for the Toy Model of Abundance Ratios\tablenotemark{a} 
Applied to the Ursa Minor Giants \label{table_fit}}
\tablehead{\colhead{Species [X/Fe]} & 
\colhead{A(X)} & \colhead{B(X)} & \colhead{[Fe/H](A)} &
\colhead{[Fe/H](B)} & \colhead{[Fe/H](low,X)} &  
\colhead{[Fe/H](high,X)}  \\
\colhead{} & \colhead{(dex)} & \colhead{(dex)} & \colhead{(dex)} & \colhead{(dex)}
 & \colhead{(dex)} & \colhead{(dex)} }
\startdata    
${\rm{[Na/Fe]}}$ & 
   $-0.33$ & $-0.70$ & $-3.09$ & $-1.58$ & $-2.18$ & $-1.75$ \\
${\rm{[Mg/Fe]}}\tablenotemark{c}$ & 
      0.36 & 0.22 & $-3.09$ & $-1.58$ & $-1.94$ & $-1.82$ \\
~~thick disk\tablenotemark{b} &
   0.35 & 0.00 & $-1.00$ & 0.00 & $-0.53$ & $-0.07$ \\     
~~thin disk\tablenotemark{b} &
     0.11 & 0.02 & $-$0.70 & 0.05 & $-0.50$ & $-0.10$ \\
${\rm{[Si/Fe]}}$ &
    0.67 & 0.05 & $-3.09$ & $-1.58$ & $-2.91$ & $-1.64$  \\
${\rm{[Ca/Fe]}}$ &
    0.19 & 0.04 & $-3.09$ & $-1.58$ & $-2.67$ & $-2.12$ \\
${\rm{[Sc/Fe]}}$ &
    $-0.13$ & $-0.17$ & $-3.09$ & $-1.58$ & $-1.88$ & $-1.82$ \\
${\rm{[Ti12/Fe]}}\tablenotemark{c}$ & 
  0.13 & 0.04 & $-3.09$ & $-1.58$ & $-3.03$ & $-1.64$ \\
\enddata
\tablenotetext{a}{The model and its parameters are described 
in \S\ref{section_toy_model}.  The low outlier COS171 is not used
for any of the fits.}
\tablenotetext{b}{Fits to the Milky Way thin and thick disk sample
of \cite{reddy03} and \cite{reddy06} for [Mg/Fe] vs [Fe/H].}
\tablenotetext{c}{${\rm{[Ti12/Fe]}} ~ = ~{\rm{[Ti ~ from ~ TiI/Fe~ from ~ FeI]}} 
~ + ~ {\rm{[Ti ~ from ~ TiII/Fe ~ from ~ FeII]}})/2$.}
\tablenotetext{d}{An offset of +0.2~dex was applied to [Mn/Fe] for the two
lowest [Fe/H] stars as only the 4030~\AA\ lines were detected.}
\tablenotetext{e}{$\alpha$ elements =  Mg, Si, Ca and Ti12.
Fe-peak elements = Cr, Mn, and Ni. 2-$\sigma$ uncertainties are tabulated.}
\tablenotetext{f}{The values of A(X), B(X), [Fe/H](A) and [Fe/H](B) were 
assumed from each individual fit.}
\tablenotetext{g}{``Combo'' combines Si, Cr, Mn, Ni, and
Ba,  the elements with the strongest dependence
of [X/Fe] on [Fe/H] in our UMi sample.  2-$\sigma$ uncertainties are tabulated.}
\end{deluxetable}


\clearpage

\clearpage

\clearpage

\clearpage

\clearpage

\clearpage

\clearpage

\end{document}